# REACTION RATE OF $p^{14}\text{N} \to {}^{15}\text{O}\gamma$ CAPTURE TO ALL BOUND STATES IN POTENTIAL CLUSTER MODEL

Sergey Dubovichenko[*,1,2,‡], Nataliya Burkova[2,¶], Albert Dzhazairov-Kakhramanov[*,1,§] and Bekmurza Beysenov[1,2]

[1] *Fesenkov Astrophysical Institute "NCSRT" ASA MDASI Republic of Kazakhstan (RK) Observatory 23, Kamenskoe plato, 050020, Almaty, RK*
[2] *Al-Farabi Kazakh National University, 050040, av. Al-Farabi 71, Almaty, RK*
[‡]*dubovichenko@mail.ru*, [§]*albert-j@yandex.ru*, [¶]*natali.burkova@gmail.com*

**Abstract:** Review of calculation results for astrophysical *S*-factor of the ${}^{14}\text{N}(p, \gamma){}^{15}\text{O}$ capture reaction in the $p^{14}\text{N}$ channel of ${}^{15}\text{O}$ was presented. It was carried out in the frame of the modified potential cluster model, taking into account resonances in the ${}^{15}\text{O}$ spectrum up to 3.2 MeV at energy of incident protons varying from 30 keV to 5 MeV. It is possible to describe experimental data for the astrophysical *S*-factors of the radiative proton capture on ${}^{14}\text{N}$ to five excited states of ${}^{15}\text{O}$ at excitation energies from 5.18 MeV to 6.86 MeV, only under assumption, that all five resonances are *D* scattering waves. Quality new physical interpretation of the capture mechanism is discussed in this channel to the ground state of ${}^{15}\text{O}$. Carried out by us assumption that the ground state of ${}^{15}\text{O}$ is determined by the $p^{14}\text{N}^*$ channel with excited ${}^{14}\text{N}^*$ cluster, immediately allowed us to correctly describe order of values of the experimental *S*-factor for capture to this state. Taking into account these results, the total *S*-factor of the proton capture on ${}^{14}\text{N}$ and the reaction rates to the ground and five excited states of ${}^{15}\text{O}$ were determined at temperatures from 0.01 to 10 $T_9$. The parametrization of the total reaction rate with a simple form is performed, which allows to obtain $\chi^2$ equal to 0.06 with 5% errors of the calculated rate.

## 1. Introduction

Continuing to study processes of radioactive capture, we consider $p^{14}\text{N} \to {}^{15}\text{O}\gamma$ capture in framework of a modified potential cluster model (MPCM) with forbidden states (FS),[1,2] which effectively take into account Pauli exclusion principle.[3] First of all, the process with capture to the fourth excited state (4[th]ES) of ${}^{15}\text{O}$ at an excitation energy of 6.79 MeV is considered, since it makes the greatest contribution to total *S*-factor. The astrophysical *S*-factor of capture to the 4[th]ES at low energies has the largest value compared to capture to other bound states (BS), reaching, apparently, up to 70–80% of total *S*-factor of capture. This is a reason for primary consideration of the *S*-factor of the capture to the 4[th]ES, and not to the ground state (GS) of ${}^{15}\text{O}$, as usually we considered other capture reactions.[1,2]

In this work we present materials earlier partially published in Refs. 4–8. Furthermore, we consider $p^{14}\text{N} \to {}^{15}\text{O}\gamma$ capture reaction to all other excited states (ES) of ${}^{15}\text{O}$. It is shown that experimental data for the astrophysical *S*-factor for capture to all excited states can be described only under assumption that all five considered resonances at energies from 260 keV to 3.2 MeV are in ${}^4D_{1/2}$ and ${}^{2+4}D_{3/2}$ scattering waves. Notations ${}^{2+4}D_{3/2}$ or ${}^{2+4}P_{1/2}$, i.e. ${}^{(2S_1+1)+(2S_2+1)}L_J$, determine mixed with possible spin values of channel $S_1$ and $S_2$ state with a given total *J* and orbital *L* moments.

---
[*] Corresponding authors: albert-j@yandex.ru, dubovichenko@mail.ru

In conclusion, we consider capture reaction with transition to the ground state of $^{15}$O, under assumption that it is a mixture of $^{2+4}P_{1/2}$ states. We note that quantitative agreement with available experimental data for the $p^{14}$N → $^{15}$Oγ radiative capture process to the GS, using $^{2+4}P_{1/2}$ waves, have not been achieved. After reviewing quite a lot of different works, we came to conclusion that no one had yet succeeded in describing these data when capturing to the GS based on any model calculations.

In particular, known to us attempt to find a qualitative explanation of failures in describing proton capture on $^{14}$N to the GS of $^{15}$O was made earlier in Ref. 9 and carried out on basis of the shell model (SM). Apparently, this is only work where "problem" of ground state is discussed within framework of structural features of SM. In intermediate coupling model, wave function (WF) of the GS is written as a superposition of two components $A \cdot [^{14}N(1_1^+) \otimes p_{3/2}] + B \cdot [^{14}N(1_1^+) \otimes p_{1/2}]$ whose amplitudes can be varied. However, as authors point out, root causes for determining absolute values of weighting factors A and B, which define corresponding cross sections, have not been found until now. As a result, it is also not possible to correctly describe experimental *S*-factor capture to the GS of $^{15}$O in this work, although all the known data for capture to the ES are described quite accurately.

In addition, Ref. 10 presents results of calculations of 34 processes, both radiative capture of protons and other reactions at astrophysical energies in a potential model (PM). This includes $p^{14}$N → $^{15}$Oγ capture process to the GS and all excited $^{15}$O states. It seems to us that these calculations are essentially fit for experimental data, since each of the partial cross sections is normalized to its value at resonance, and then it is proposed to consider such cross section as reliable. Our comments, however, should not be taken as a criticism of these results, but only as brief explanations of the work done in Ref. 10, which can also be useful, especially at this stage of consideration of the reaction.

There are two more works, in the first of which only proton capture on $^{14}$N to the 4$^{th}$ES of $^{15}$O was considered within framework of potential model,[11] as well as Ref. 12, in which the same capture to 4$^{th}$ES, and the potentials between the particles of 16 reactions considered were based on the convolution model of M3Y potentials.[13] However, the capture to the GS in these papers was not considered at all.

As a result, we were unable to find any theoretical calculations within, for example, RGM or "*ab initio*" methods for astrophysical characteristics of this reaction. It seems that almost all of available works, with exception of the above, is limited to only *R*-matrix processing of existing experimental data. And since the quality of this data leaves much to be desired, results of *R*-matrix analysis, which uses dozens of fitting parameters, performed in different works and different years, differ greatly among themselves (see, for example, Refs. 14,15 and references in them to earlier works).

## 1.1. *Physical aspects of the used method*

The presence of an FS in a two-particle single-channel MPCM is determined on the basis of classification of orbital states of clusters according to Young diagrams.[3] In our approach potentials of intercluster interactions for scattering processes are based on the reproduction of elastic scattering phase shifts of considered particles with their resonance behavior or based on structure of spectra of resonance states of final nucleus. For BS of nuclei formed as a result of capture reaction in cluster channel, which coincides with the initial particles, intercluster potentials are constructed based on the description of binding energy of these particles in final nucleus and some basic characteristics of such states (see, for example, Refs. 16–18). Usually, these are charge radius and asymptotic constant (AC), meanwhile, the characteristics of clusters connected in nucleus are identified with characteristics of the correspondent free nuclei.[19]

The single-channel and two-body character of model is obviously idealization, since it assumes that in the BS nucleus has 100% clusterization in a certain two-particle channel. Therefore, success of this cluster model in the description of system of A nucleons in a bound state is determined by the fact that the BS clusterization of such nucleus in the two-particle channels is large.[20] However, some nuclear characteristics of individual, even non-cluster nuclei, in certain reactions may be caused mainly by one distinct cluster channel, with little contribution from other possible cluster configurations. In this case, the used single-channel two-cluster model makes it possible to identify the dominant cluster channel and to highlight some of the basic properties of such a nuclear system that are determined by them.[1,2,19]

**1.2.** *Astrophysical aspects of the reaction*

The considered proton capture reaction on $^{14}$N to various BS of $^{15}$O is included in CNO cycle of thermonuclear processes, which largely determine energy release processes of our Sun and, possibly, many stars of our Galaxy. Apparently, a fairly modern description of astrophysical applications associated with the reaction of radiative proton capture on $^{14}$N is given in review paper.[21]

In particular, it is shown that globular clusters are the oldest stellar formations. Their age coincides with time of first star's formations in the Universe and provides an independent check of reliability of standard (and non-standard) cosmological models. The main sequence stars currently observed in globular clusters have masses smaller than mass of the Sun. These low-mass stars burn in their center, mainly due to the thermonuclear *pp*-chain.

However, at the end of their lifetime, when central fraction of hydrogen mass becomes less than about 0.1, nuclear energy emitted by H-burning becomes insufficient and core of the star contracts, which leads to energy extraction from its own gravitational field. Then, central temperature and density increase, and burning process of the star switches from the *pp*-chain to a more efficient CNO-cycle. However, bottleneck of CNO cycle is the $^{14}$N($p,\gamma$)$^{15}$O reaction considered here – radiative capture of proton on $^{14}$N. There was a long-standing problem between observations and stellar models. It has been long observed overproduction of all chemical elements above carbon by a factor of 2 or more. This discrepancy was resolved by using a lower reaction rate $^{14}$N($p,\gamma$)$^{15}$O for evolution of chemical elements.[21]

So, it can be considered that the $^{14}$N($p,\gamma$)$^{15}$O reaction considered here is of great importance for various tasks and study of its mechanism plays an important role for nuclear physics, since so far it has not been possible to describe capture on the GS in the frame of certain model. Such a reaction, as it was shown above, plays an important role for nuclear astrophysics and its consideration in framework of nuclear physics methods should make it possible to determine real form and magnitude of the reaction rate.

## 2. Analysis of cluster states of N+$^{14}$N system

Turning to analysis of cluster states of $N+^{14}$N system with formation of $^{15}$N or $^{15}$O, we note that classification of orbital states of $^{14}$C in the $n^{13}$C system or $^{14}$N in the $p^{13}$C channel according to Young diagrams were considered by us earlier in Ref. 22. However, since we do not have total tables of Young diagrams products for system with more than eight particles,[23] which we used earlier for similar calculations,[1,2] results obtained further should be considered only a qualitative assessment of possible orbital symmetries in the BS of $^{15}$O and $^{15}$N nuclei for channels under consideration.

### 2.1. *State classification by Young diagrams*

At the same time, based on such a classification it was quite acceptable to explain available experimental data, for example, on radiative capture for $N^2$H Refs. 17,22, $N^3$H Ref. 18, $n^{12}$C Refs. 2,22, $p^{12}$C Refs. 22,24, $n^{13}$C Refs. 2,22, $p^{13}$C Refs. 22,25 or $n^{14}$C and $n^{14}$N Refs. 2,22 systems. Therefore, here we will use the classification of cluster states according to orbital symmetries, which leads to a certain number of bound FS and allowed states (AS) in partial intercluster potentials, and, therefore, to a certain number of nodes for relative cluster motion wave function.

Furthermore, we assume that for $^{14}$A nucleus, we can accept the Young orbital diagram in the form {4442}, therefore for the $N+^{14}$A system in framework of $1p$-shell we have {1} × {4442} → {5442} + {4443}.[20,23] The first diagram is compatible with orbital momenta $L = 0$ and 2, and is forbidden, since there cannot be five nucleons in $s$-shell,[3,20] and the second one is allowed and compatible with orbital momentum $L = 1$.

Thus, limited with only the lowest partial waves with orbital momenta $L = 0, 1, 2$, we can say that for the $n^{14}$C system[22] (for $^{14}$C we have $J^\pi, T = 0^+1$ Ref. 26) in potential $^2S_{1/2}$ wave there are forbidden and allowed state. The last of them corresponds to the GS of $^{15}$C with $J^\pi = 1/2^+$ and is located at binding energy of the $n^{14}$C system -1.21809 MeV.[28] The potentials of the $^2P$ waves of elastic scattering do not have the FS, and the bound FS may be present in the $^2D$ waves.[22]

For the $p^{14}$C system,[22,25] potential of the $S$-scattering wave contains a forbidden bound state, and the $P_{1/2}$ wave has an allowed bound state, which corresponds to the GS of $^{15}$N with $J^\pi = 1/2^-$ and is at binding energy of the $p^{14}$C system, equals -10.2074 MeV.[26]

In case of the $n^{14}$N system[22] (for $^{14}$N we have $J^\pi, T = 1^+0$ Ref. 26), forbidden bound state is present in potentials of the $S$ scattering waves, and the $P_{1/2}$ wave has only allowed state, which corresponds to the GS of $^{15}$N с $J^\pi = 1/2^-$, and is found at binding energy of the $n^{14}$N system, equal to -10.8333 MeV.[26]

We find similar situation for the $p^{14}$N system – in potentials of the $S$ and $D$ scattering waves there is a forbidden bound state, and the $P_{1/2}$ wave has only the AS, which corresponds to the GS of $^{15}$O with $J^\pi = 1/2^-$ and is at the binding energy of the $p^{14}$N systems equal to -7.2971 MeV.[26] It is possible to match spin-mixed $^2P_{1/2}+^4P_{1/2}$ states, which lead to the same total moment, with this ground state of $^{15}$O with the $^{14}$N cluster in its GS. The question of possibility of the $^{14}$N cluster excitation and using the $D_{1/2}$ wave of $^{15}$O nucleus for the GS are discussed further.

## 2.2. Structure of excited and resonance states in $^{15}$O

Now we give information about excited states in $^{15}$O, which are below threshold of the $p^{14}$N channel, in order to present a general picture of spectrum of bound levels, which is shown in Fig. 1a

1. With an excitation energy of 5.183(1) MeV above BS or -2.1141 MeV in c.m. Ref. 26 relative to threshold of the $p^{14}$N channel, there is a first excited, but bound in this channel, state with the moment $J^\pi = 1/2^+$, which can be matched to the doublet $^2S_{1/2}$ wave with FS. Although such state can, of course, be considered as a quartet $^4D_{1/2}$ wave with FS.

2. The second ES (2$^{nd}$ES) with an excitation energy of 5.2409(3) MeV Ref. 26 relative to the GS or -2.0562 MeV relative to the threshold of the $p^{14}$N channel has $J^\pi = 5/2^+$ and it can be matched to the mixture of doublet and quartet $^{2+4}D_{5/2}$ waves with FS.

3. The third excited state (3$^{rd}$ES) at an excitation energy of 6.1763(17) MeV Ref. 26 relative to the GS or -1.1208 MeV relative to channel threshold has $J^\pi = 3/2^-$ and it can be compared to a mixture of doublet and quartet $^{2+4}P_{3/2}$-waves without an associated FS.

4. The fourth excited state (4$^{th}$ES) has an excitation energy of 6.7931(17) MeV relative to the GS or –0.504 MeV Ref. 26 or relative to the channel threshold has $J^\pi = 3/2^+$ and it can be compared with the quartet $^4S_{3/2}$ wave with the FS. Of course, this state may also be a mixture of doublet and quartet $^{2+4}D_{3/2}$ waves with the FS.

5. The fifth excited state (5$^{th}$ES) with an excitation energy of 6.8594(9) MeV Ref. 26 relative to the GS or -0.4377 MeV relative to the channel threshold has $J^\pi = 5/2^+$ and it can also be compared to a mixture of doublet and quartet $^{2+4}D_{5/2}$ wave with the FS

6. There is another bound state in the $p^{14}$N channel with $J^\pi = 7/2^+$ at an excitation energy of 7.2759(6) MeV Ref. 26 relative to the GS or -0.0212 MeV relative to the threshold of the considered channel, which can be compared to the $^4D_{7/2}$ state of the $p^{14}$N system. However, we will not consider it because of too small binding energy and large total moment. Indeed, from the results, for example, of Ref. 27, it is clear that this level makes the smallest contribution to the total *S*-factor of the proton capture on $^{14}$N, and the maximum contribution comes from the capture to the 4$^{th}$ES and the GS of $^{15}$O. For this reason, the level with $J^\pi = 7/2^+$ is not shown in Fig. 1a.

```
10.506 (3/2⁻)
─────────────
9.609(2) (3/2⁻)
─────────────
9.484(8) (3/2⁺)
─────────────
8.743(6) (1/2⁺)
─────────────
8.2840(5) (3/2⁺)
─────────────
7.5565(4) (1/2⁺)    p¹⁴N
                    7.2971
─────────────
6.8594(9) (5/2⁺)
─────────────
6.7931(17) (3/2⁺)
─────────────
6.1763(17) (3/2⁻)
─────────────
5.2409(3) (5/2⁺)
─────────────
5.183(1) (1/2⁺)
─────────────

¹⁵O (1/2⁻,1/2)
```

Fig. 1a. Level structure of $^{15}$O.

Furthermore, we consider the spectrum of some resonance states in the $p^{14}$N system (see Fig. 1a), i.e., states at positive energies above the threshold of the $p^{14}$N channel

1. The first resonance state (1$^{st}$RS) of $^{15}$O in the $p^{14}$N channel is at an excitation energy of 7.5565(4) MeV or 259.4(4) keV in c.m. relative to the threshold of the $p^{14}$N channel (see Table 15.16 or 15.20 in Ref. 26), has a width of 0.99(10) keV in c.m. and moment $J^\pi = 1/2^+$. Such resonance can be matched with the $^2S_{1/2}$ wave of $p^{14}$N scattering with the FS, although, of course, this state can correspond to the quartet $^4D_{1/2}$ resonance wave with the FS.

2. The second resonance state ($2^{nd}$RS) has an excitation energy of 8.2840(5) MeV or 986.9(5) keV c.m. relative to channel the threshold, its width is equal to 3.6(7) keV in c.m. and the moment $J^\pi = 3/2^+$.[26] Therefore, it can be matched with $^4S_{3/2}$ scattering wave with FS. This state, generally speaking, can be a mixed doublet and quartet $^{2+4}D_{3/2}$ resonance scattering wave.

3. The third resonance state ($3^{rd}$RS) has an excitation energy of 8.743(6) MeV or 1446(6) keV c.m. relative to the threshold, its width is 32 keV in c.m. and the moment $J^\pi = 1/2^+$.[26] Therefore, it can be matched with a $^2S_{1/2}$ scattering wave with the FS or with the quartet $^4D_{1/2}$ resonance scattering wave.

4. The fourth resonance state ($4^{th}$RS) has an excitation energy of 9.484(8) MeV or 2187(8) keV c.m. relative to the threshold of channel, its width is equal to 191 keV in c.m. and the moment, apparently, has the value $J^\pi = 3/2^+$.[26] Therefore, it can be matched with the $^4S_{3/2}$ scattering wave with the FS or with the mixed doublet and quartet $^{2+4}D_{3/2}$ resonance scattering waves.

5. The fifth resonance state ($5^{th}$RS) has an excitation energy of 9.609(2) MeV or 2312(2) keV c.m. relative to the channel threshold, its width is equal to 8.8(5) keV in c.m. and the moment $J^\pi = 3/2^-$.[26] Therefore, it can be compared with the $^4P_{3/2}$ or $^4F_{3/2}$ scattering waves without the FS.

6. The sixth resonance state ($6^{th}$RS) has an excitation energy of 10.506 MeV or 3209 keV c.m. relative to the channel threshold, its width is equal to 140(40) keV in c.m. and the moment $J^\pi = 3/2^+$.[26] Therefore, it can be compared with the $^4S_{3/2}$ scattering wave with the FS or with the mixed doublet and quartet $^{2+4}D_{3/2}$ resonance scattering wave.

We did not consider some states and did not list them in Fig. 1a, if they have not well defined moment.[26] Levels of small width with high excitation energy (greater than 1 MeV), as well as states at energies above 3.2 MeV, were not considered. All these resonances in experimental data of the proton capture on $^{14}$N used by us are not observed. One of the reasons is due to the large energy step in measurements of total cross sections.

At energies up to about 3.5 MeV, there are no resonance levels of wide width that could be compared with doublet or quartet $P$ scattering waves Refs. 26 and 28 in spectra of $^{15}$O. Therefore, phase shifts of these partial waves can be taken equal to or close to zero, and since there are no bound forbidden states in the $P$ waves, the depth of these potentials as the first option (the second is considered further), can simply equal to zero.

## 3. Calculation method

Astrophysical $S$-factors characterize behavior of the total cross section for nuclear reaction when energy tends to zero, and are defined as follows[29]

$$S(NJ, J_f) = \sigma(NJ, J_f) E_{cm} \exp\left(\frac{31.335 Z_1 Z_2 \sqrt{\mu}}{\sqrt{E_{cm}}}\right), \quad (1)$$

where $\sigma(NJ,J_f)$ is the total cross section of the radiative capture process in barn, $E_{cm}$ is a particle energy, usually measured in keV, in center of mass system, $\mu$ is a reduced mass of particles in initial channel in amu, $Z_{1,2}$ are particle charges in units of an elementary charge and $N$ is $E$ or $M$ transitions of the $J^{th}$ multipolarity to final $J_f$ state of the nucleus. The value of numerical coefficient 31.335 is obtained on basis of modern values of fundamental constants.[30]

The total cross sections of the radiative capture $\sigma(NJ,J_f)$ for $EJ$ and $M1$ transitions in the potential cluster model are given, for example, in Ref. 31 or Refs. 1,2,32,33 and have the form

$$\sigma_c(NJ,J_f) = \frac{8\pi K e^2}{\hbar^2 q^3} \frac{\mu}{(2S_1+1)(2S_2+1)} \frac{J+1}{J[(2J+1)!!]^2} A_J^2(NJ,K) \sum_{L_i,J_i} P_J^2(NJ,J_f,J_i) I_J^2(J_f,J_i)$$

where $q$ is the wave number of particles in initial channel, $S_1$, $S_2$ are the spins of particles in initial channel, $K$, $J$ are wave number and moment of $\gamma$ quantum in final channel.

For electric orbital $EJ(L)$ transitions ($S_i = S_f = S$), coefficients in this expression have the form Refs. 1,2,17

$$P_J^2(EJ,J_f,J_i) = \delta_{S_i S_f} \left[(2J+1)(2L_i+1)(2J_i+1)(2J_f+1)\right](L_i 0 J 0 | L_f 0)^2 \begin{Bmatrix} L_i & S & J_i \\ J_f & J & L_f \end{Bmatrix}^2,$$

$$A_J(EJ,K) = K^J \mu^J \left(\frac{Z_1}{m_1^J} + (-1)^J \frac{Z_2}{m_2^J}\right), \qquad I_J(J_f,J_i) = \langle \chi_f | r^J | \chi_i \rangle.$$

Here $S_i$, $S_f$, $L_f$, $L_i$, $J_f$, $J_i$ are total spins and moments of particles of initial ($i$) and final ($f$) channels, $m_1$, $m_2$ are particle masses of initial channel, $I_J$ – integral of the wave functions for initial $\chi_i$ and final $\chi_f$ state, as functions of relative motion of clusters with intercluster distance $r$.

For consideration of the magnetic $M1(S)$ transition, caused by spin part of magnetic operator,[34] using expression,[35] one can get Refs. 1,2,17

$$P_1^2(M1,J_f,J_i) = \delta_{S_i S_f} \delta_{L_i L_f} \left[S(S+1)(2S+1)(2J_i+1)(2J_f+1)\right] \begin{Bmatrix} S & L & J_i \\ J_f & 1 & S \end{Bmatrix}^2,$$

$$A_1(M1,K) = i\frac{\hbar K}{m_0 c}\sqrt{3}\left[\mu_1 \frac{m_2}{m} - \mu_2 \frac{m_1}{m}\right], \qquad I_J(J_f,J_i) = \langle \chi_f | r^{J-1} | \chi_i \rangle, \quad J=1,$$

where $m$ is the total mass of nucleus, $\mu_1$ and $\mu_2$ are magnetic moments of clusters.

If the total reaction cross sections or astrophysical $S$-factor of the radiative capture process are known, the rate of this reaction, which is used for astrophysical applications, can be determined as Ref. 31

$$N_A \langle \sigma v \rangle = 3.7313 \cdot 10^4 \mu^{-1/2} T_9^{-3/2} \int_0^\infty \sigma(E) E \exp(-11.605 E/T_9) dE,$$

where $E$ is the energy specified in MeV, cross section $\sigma(E)$ is measured in μb, μ is the reduced mass in amu, $T_9$ is the temperature in $10^9$ K.

In present calculations, the following values of particle masses $m_p$ = 1.007276469 amu[36] and $m(^{14}N)$ = 14.003074 amu were used,[37] constant $\hbar^2/m_0$ was taken to be 41.4686 MeV·fm$^2$, $m_0$ is an atomic unit of mass. The Coulomb parameter $\eta = \mu Z_1 Z_2 e^2/(q\hbar^2)$ was represented as $\eta = 3.44476 \cdot 10^{-2} \cdot Z_1 Z_2 \mu/q$, where $q$ is the wave number expressed in fm$^{-1}$, determined by energy of interacting particles in initial channel. The Coulomb potential for zero Coulomb radius $R_{coul}$ = 0 is written in form $V_{coul}(MeV) = 1.439975 \cdot Z_1 Z_2 / r$, where $r$ is relative distance between particles of initial channel, expressed in fm. Magnetic moments are $\mu_p$ = 2.792847$\mu_0$ Ref. 36 and $\mu(^{14}N)$ = 0.404$\mu_0$ Ref. 38, where $\mu_0$ is a nuclear magneton.

### 3.1. *Criteria for constructing the interaction potentials*

Now let us dwell in more detail on procedure for constructing intercluster partial potentials used here for a given orbital moment $L$, channel spin $S$ and total moment $J$, by defining criteria and sequence for finding parameters and indicating their errors and ambiguities.[1,2] First of all, we find parameters of BS potentials, which, for a given number of allowed and forbidden bound states in a given partial wave $^{(2S+1)}L_J$, are fixed quite uniquely by binding energy, radius of nucleus and AC in the considering channel. The accuracy, with which parameters of the BS potential are determined, is associated primarily with accuracy of AC, which is usually 10–20%. This potential does not contain other ambiguities, since classification of states according to Young diagrams makes it possible to uniquely fix number of BSs that are forbidden or allowed in a given partial wave, which fully determines its depth, and width of potential completely depends on AC value.

The intercluster potential of nonresonance scattering process by scattering phase shifts for a given number of BSs allowed and forbidden in considered partial wave, is also constructed quite clearly. The accuracy of determining parameters of such potential is connected, first of all, with the accuracy of scattering phases from experimental data and can reach 20–30%. Here, such a potential does not contain ambiguities, since classification of states according to Young diagrams makes it possible to fix unambiguously the BS number, which totally determines its depth, and potential width at the given depth is determined by the shape of scattering phase shift.

Constructing the nonresonance scattering potential from data on nuclear spectra in a certain channel, it is difficult to estimate accuracy of finding its parameters, even with a given number of BSs, although it can be expected that it slightly exceeds the error in previous case. Such a potential, as it is usually assumed for energy range up to 1 MeV, should lead to the scattering phase shift close to zero or give a smooth falling form of phase shift, if there are no resonance levels in spectra of the nucleus.

In process of analyzing resonance scattering, when a comparatively narrow resonance is present in the considered partial wave, then for a given number of BSs, potential is totally uniquely constructed. For a given number of BSs, its depth is uniquely fixed by resonance energy of level, and width is totally determined by width of such a resonance. The error of its parameters usually does not exceed error of determining width of this level and is approximately 3–5%. Moreover, this also applies to construction of partial potential from scattering phase shifts and determination of its parameters from resonance in the spectra of the nucleus.

As a result, all intercluster potentials do not contain ambiguities, to which the optical model[39] leads, and allow, in general, one to correctly describe total cross sections of processes of radiative capture. Previously, we considered more than 30 such reactions[1,2,17,22] and in all cases the described criteria for constructing potentials fully justified themselves.

We emphasize once again that in construction of partial interaction potentials, it is believed that they depend not only on orbital moment $L$, but also on total spin $S$ and total moment $J$ of cluster system. In other words, for different moments of $JLS$, we will have different parameter values. Since $E1$ transitions with a change in parity between different $^{(2S+1)}L_J$ states in continuous and discrete spectra are usually considered, the potentials of these states are different.

In addition, one of modifications of the model we use is an assumption that the intercluster potentials are explicitly dependent on Young diagrams.[3] Here it is possible that scattering states and BS have a different number of Young diagrams. In other words, if two diagrams are allowed in the continuous spectrum states, and only one in discrete spectrum, such potentials can have different parameters with the same $JLS$, i.e., in the same partial wave.[1–3] Therefore, when considering $M1$ transitions, there are cases when a transition occurs between states with the same $JLS$, but different $\{f\}$.[2,3] We will dwell on these issues in more detail when considering each capture process for a specific BS.

As a result, when calculating astrophysical $S$-factors of the proton capture on $^{14}$N, we consider two variants of the $p^{14}$N system potentials. First, based on description of characteristics of scattering processes, partial potentials of continuous spectrum are constructed (see Table 1). These characteristics primarily include parameters of resonances located above threshold of the $p^{14}$N channel in $^{15}$O, i.e., in continuous spectrum. Then, based on description of binding energy and asymptotic constant for each BS, partial potentials of these states are obtained (see Table 2).

These interaction potentials, obtained independently from one another, are used in the first version of calculations. In the second version of calculations, interactions of BSs in the corresponding partial waves are used as scattering potentials. However, this is possible only if there is no resonance of scattering in such a partial wave. It is not possible to construct a Gaussian type potential that has resonance for continuous spectrum and BS with correct characteristics for discrete spectrum. Therefore, in this case, different partial potentials is used for scattering process, which has a resonance and describes its characteristics, and a bound state, which is obtained on basis of description of binding energy and ACs of this BS.

It is known that potential of two structureless point particles considered by quantum mechanics must be the same in continuous and discrete spectrum. However, system of nuclear particles is not structureless and even more it is not pointed. At least, therefore, question of "uniformity" of potentials of a complex nuclear system in a continuous and discrete spectrum cannot be considered as unambiguously solved.

### 3.2. Potentials of the $p^{14}$N interaction in different states

We use a simple Gaussian potential of the form as in our previous works Refs. 1,2,16–18,22, for description partial intercluster interaction (in this case between a nucleon and a nucleus),

$$V_{JLS}(r) = -V_{0,JLS}(r) \cdot \exp(-\alpha_{JLS} r^2) + V_{Coul}(r)$$

with a point-like Coulomb term, the form of which is given above. Let us now consider the partial potentials for the resonance and excited states of the $p^{14}N$ system in $^{15}O$.

As already mentioned, in processes of the elastic $p^{14}N$ scattering and spectrum of $^{15}O$, the first resonance is at an energy of about 260 keV in c.m. above the threshold of the $p^{14}N$ channel and has a moment $J^\pi = 1/2^+$.[26] As a first option, it can be matched with resonance $^2S_{1/2}$ wave of the $p^{14}N$ scattering. Then, for a potential with the FS constructed on the basis of spectra of final nucleus,[26] the parameters were obtained, which are listed in Table 1 at number 1. With this potential, the resonance energy of level 259.4(1) keV in c.m. is found with a width of 0.98(1) keV in c.m., which agrees well with experimental data.[26] For this energy, the scattering phase shift, if we do not take into account generalized Levinson theorem, mentioned below, turned out to be 90.0°(1). The shape of such a resonance $^2S_{1/2}$ scattering phase shift is shown in Fig. 1b by the red solid curve. We used a well-known expression $\Gamma = 2(d\delta/dE)^{-1}$ for calculation the level width $\Gamma$ using scattering phase shift $\delta$, depending on the energy $E$.

Table 1. Potential parameters of the $p^{14}N$ scattering in different partial waves.

| No. | Scattering State | $V_0$, MeV | $\alpha$, fm$^{-2}$ | Energy of the resonance in c.m., keV | Width of the resonance $\Gamma_{cm}$, keV |
|---|---|---|---|---|---|
| 1 | Resonance in $^2S_{1/2}$ | 93.29725 | 0.17 | 259.4(1) | 0.98(1) |
| 2 | 1$^{st}$RS in $^4D_{1/2}$ | 20.029545 | 0.013 | 260.0(1) | 0.97(1) |
| 3 | Third resonance by turn in spectrum or 2$^{nd}$RS in $^4D_{1/2}$ | 571.622 | 0.25 | 1447(1) | 33.5(1) |
| 4 | Second resonance by turn in spectrum or 1$^{st}$RS in $^{2+4}D_{3/2}$ | 915.61367 | 0.4 | 987.0(1) | 3.1(1) |
| 5 | Fourth resonance by turn in spectrum or 2$^{nd}$RS in $^{2+4}D_{3/2}$ | 474.42 | 0.21 | 2187(1) | 197(1) |
| 6 | Sixth resonance by turn in spectrum or 3$^{rd}$RS in $^{2+4}D_{3/2}$ | 1463.247 | 0.65 | 3211(1) | 131(1) |
| 7 | Nonresonance $^2S_{1/2}$ and $^4S_{3/2}$ | 3500.0 | 10.0 | --- | --- |
| 8 | Nonresonance $^{2+4}D_{5/2}$ and $^4D_{7/2}$ | 1000.0 | 1.0 | --- | --- |
| 9 | Nonresonance $^2P$ and $^4P$ | 0.0 | 1.0 | --- | --- |
| 10 | Fifth resonance by turn in spectrum in $^4F_{3/2}$ | 171.04523 | 0.16 | 2312(1) | 9.0(1) |

If we take into account the generalized Levinson Theorem,[3,40] then if there is one bound FS, but without associated AS, the phase shifts of the $S$ and $D$ potentials No. 1–No. 8 from Table 1 should start at zero energy from 180 degrees. Therefore, it is realistic for $S$ potential No. 1 resonance, as well as for all other resonance potentials, the phase shift is equal to 270.0(1)°. Furthermore, in the text, for all $S$ and $D$ scattering waves, values of the phase shifts are given taking into account Levinson theorem. However, in Fig. 1b, scattering phase shifts for all potentials that are used in these calculations are given from zero degrees, i.e., in a more familiar form.

As a second option, the first resonance state can be attributed to the $^4D_{1/2}$ scattering wave, then its parameters are listed in Table 1 under No. 2. With this potential, the resonance energy of level is equal to 260.0(1) keV in c.m. at a width of 0.97(1) keV in c.m. – for this energy, scattering phase shift taking into account the Levinson theorem was found to be 270.0(1)°. The shape of this phase shift is shown in Fig. 1b by the black dotted curve and almost coincides with a red continuous curve.

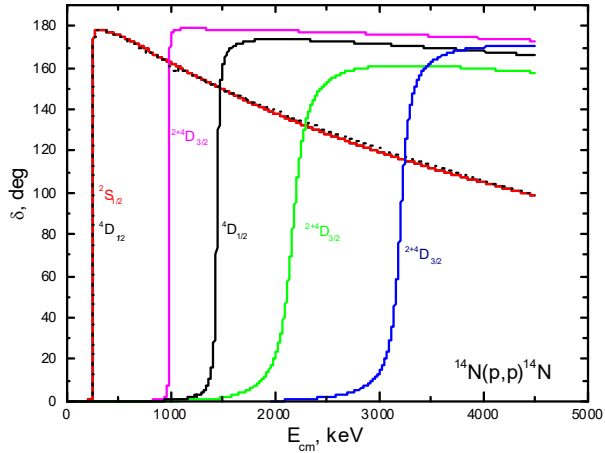

Fig. 1b. Phase shifts of the elastic $^2S_{1/2}$ or $^4D_{1/2}$ and $^{2+4}D_{3/2}$ scattering in the $p^{14}$N channel

Considering the second resonance state, we note that we did not manage to construct the potential, assuming that the resonance at 987 keV in c.m.. with $J^\pi = 3/2^+$ is in $^4S_{3/2}$ wave. Options of potentials make it possible to obtain resonance energy, but the calculated width of such resonance turns out to be much larger than observed value of 3.7 keV in experiment.[26] However, if we assume that this resonance belongs to $^{2+4}D_{3/2}$ wave of the $p^{14}$N scattering with the FS, then we can immediately find parameters shown in Table 1 under No. 4. This potential leads to the resonance energy of 987.0(1) keV with a width of 3.1(1) keV, which agrees well with experimental data,[26] and elastic scattering phase shift is shown in Fig. 1b by the solid violet curve.

Table 1 shows the parameters of other resonance potentials No. 3, No. 5, and No. 6, which are also in $D$ waves, and their phase shifts are shown in Fig. 1b, with black, green, and blue solid curves. All of these potentials were constructed in such a way as to correctly describe characteristics of corresponding resonances, i.e., their energy and width. The accuracy with which parameters of such potentials are determined, for a given number of FS $N_{FS} = 1$, depends on accuracy of determination of energy and width for such resonances.

For the 5$^{th}$ resonance at 2.312 MeV with $J^\pi = 3/2^-$ and potential No. 10 from Table 1 it is possible to describe its characteristics only at the assumption that it exists in the $F$ wave. The variant of resonance in the $P$ wave, as in the case of the second resonance by using for it $^4S_{3/2}$ wave, does not allow correctly describe its width. Affection of this resonance due to its small width we will take into account only at the capture to the GS, because only in this case there are experimental data with high resolution by energy.

Options of the $^2S$ and $^4S$ scattering waves potentials that do not contain resonances, but include bound FSs, may have, for example, the parameters shown in Table 1 under No. 7. They lead to the scattering phase shifts from 180 to 179 degrees in energy range from zero to 1 MeV. For nonresonance $D_{5/2,7/2}$ scattering waves with ES, one can use potential, which leads to smoothly falling scattering phase shift from 180 to 179 degrees at energy up to 1 MeV – they are listed in Table 1 under No. 8.

As already mentioned, at energies below about 3.5 MeV in spectra of $^{15}$O there are no resonance levels of wide width that could be matched with doublet or quartet $P$ scattering waves.[26] Therefore, phase shifts of these partial waves can be taken equal to or close to zero, and since there are no bound forbidden states in the $P$ scattering waves, then the depth of such $P$ potentials can simply be set to be equal to zero, as shown in Table 1 at No. 9.

Furthermore, we consider two-body potentials for all BSs of $^{15}$O considered here in the $p^{14}$N channel. For potential of the ground $^{2+4}P_{1/2}$ state at the binding energy of –7.2971 MeV of $^{15}$O in the $p^{14}$N channel[26] without FS, parameters were found, which are listed in Table 2 under No. 1. The binding energy of –7.29710 MeV was obtained with such a potential, which totally coincides with experimental value,[26] charge radius $<r^2>^{1/2}$ = 2.63 fm, and dimensionless asymptotic constant $C_w$ = 6.9(1) over interval 10–15 fm. The error given here is determined by averaging AC over specified distance interval. The scattering phase shift for such potential smoothly decreases from 180° to 179° (since it does not have FS, but only one bound AS) in energy range up to 1 MeV.

Table 2. Parameters of two-body $p^{14}$N potentials for bound states of $^{15}$O

| No. | Bound level | $V_0$, MeV | $\alpha$, fm$^{-2}$ | Binding energy, MeV | $C_w$ | $<r^2>^{1/2}$, fm |
|---|---|---|---|---|---|---|
| 1 | GS $^{2+4}P_{1/2}$ | 56.727435 | 0.1 | -7.29710 | 6.9(1) | 2.63 |
| 2 | 1$^{st}$ES $^{2}S_{1/2}$ | 4334.4066 | 10.0 | -2.11410 | 1.9(1) | 2.48 |
| 3 | 1$^{st}$ES $^{4}D_{1/2}$ | 938.993145 | 0.7 | -2.11410 | 0.4(1) | 2.51 |
| 4 | 2$^{nd}$ES $^{2+4}D_{5/2}$ | 349.81073 | 0.25 | -2.05620 | 1.1(1) | 2.64 |
| 5 | 3$^{rd}$ES $^{2+4}P_{3/2}$ | 149.970703 | 0.45 | -1.12080 | 1.6(1) | 2.55 |
| 6 | 3$^{rd}$ES $^{4}F_{3/2}$ | 39.792993 | 0.03 | -1.12080 | 2.3(1) | 3.22 |
| 7 | 4$^{th}$ES $^{4}S_{3/2}$ | 112.77365 | 0.2 | -0.50400 | 8.8(1) | 2.86 |
| 8 | 4$^{th}$ES $^{2+4}D_{3/2}$ | 57.154022 | 0.037 | -0.50400 | 7.1(1) | 3.48 |
| 9 | 5$^{th}$ES $^{2+4}D_{3/2}$ | 342.1775 | 0.25 | -0.43770 | 1.0(1) | 2.67 |

For radius of $^{14}$N, known value of 2.560(11) fm was used.[41] The radius of $^{15}$O, apparently, should not differ much from the radius of $^{15}$N, for which the value of 2.612(9) fm is known.[26] The proton radius was assumed to be equal to 0.8775(51) fm Ref. 37. The accuracy of the finite-difference method (FDM)[42] for calculating binding energy was set at the level of 10$^{-5}$ MeV. As it was shown earlier,[1,2,19,22,43] the accuracy of the FDM is fully confirmed by the variational method (VM) of the binding energy calculation.[42] The coincidence of results obtained by the FDM and VM sometimes reaches a value of order $10^{-8}$ MeV.[18]

The asymptotic constant $C_w$ we use in a dimensionless form is defined in Ref. 44 as follows

$$\chi_L(r) = \sqrt{2k_0}\, C_w W_{-\eta L+1/2}(2k_0 r). \quad (2)$$

Here, $W_{-\eta L+1/2}$ is the Whittaker function, $k_0$ is wave number of the considered BS.[44]

For asymptotic normalization coefficient (ANC) $A_{NC}$ for the GS of $^{15}$O in the cluster $p^{14}$N channel, average value of 7.5(8) fm$^{-1/2}$ is obtained from Refs. 27,45,46. If we take results for spectroscopic factor $S_f$ from Refs. 47,48, with an average value of 1.16(29), then according to the well-known expression[46]

$$A_{NC}^2 = S_f C^2 \quad (3)$$

for the dimensional asymptotic constant $C$, taking into account reduced error interval, we obtain 7.4(1.5) fm$^{-1/2}$. In these works, definition of dimensional AC is used.

$$\chi_L(r) = C W_{-\eta L+1/2}(2k_0 r), \quad (4)$$

which differs from the dimensionless AC $C_w$ used here by a value $\sqrt{2k_0}$ equal, in this case, to 1.07. Therefore, in the dimensionless form for the AC, we get 6.9(1.4), which agrees well with the results for the potential of the GS. In other words, potential of the GS was constructed in such a way as to correctly reproduce this value of the AC and binding energy of this level – this immediately yields correct value of the nucleus radius.

Now we present the summary tables for the spectroscopic factors $S_f$, asymptotic normalization coefficients $A_{NC}$, dimensional AC $C$ and dimensionless AC $C_w$ for five bound ESs of $^{15}$O in the $p^{14}$N channel, which are discussed later.

Table 3. Interval of the spectroscopic factors $S_f$ for bound states of $^{15}$O from Refs. 47,48

| Level $^{15}$O, MeV | $J^\pi$ | $L_j$ | From Ref. 47 according to data of $(\tau,d)$ and $(d,n)$ reactions with references to other works | From Ref. 48 according to data of $(\tau,d)$ and $(d,n)$ reactions with references to other works | Interval of values | Average on interval |
|---|---|---|---|---|---|---|
| 0 | 1/2$^-$ | $p_{1/2}$ | 0.87–1.08 | 1.0–1.45 | 0.87–1.45 | 1.16(29) |
| 5.18 | 1/2$^+$ | $s_{1/2}$ | 0.0–0.02 | <0.007 & <0.01 | 0.00–0.02 | 0.01(1) |
| 5.24 | 5/2$^+$ | $d_{3/2}$ | 0.03–0.11 | 0.04–0.06 | 0.03–0.11 | 0.07(4) |
|  |  | $d_{5/2}$ | 0.03–0.06 |  |  |  |
| 6.17 | 3/2$^-$ | $p_{1/2}$ | 0.04–0.16 | 0.05–0.16 | 0.04–0.16 | 0.10(6) |
| 6.79 | 3/2$^+$ | $s_{1/2}$ | 0.27–0.47 | 0.39–0.56 | 0.27–0.56 | 0.41(14) |
|  |  | $d$ | <0.3 & <0.5 | --- |  |  |
| 6.86 | 5/2$^+$ | $d_{3/2}$ | 0.36–0.58 | 0.36–0.52 | 0.36–0.64 | 0.50(14) |
|  |  | $d_{5/2}$ | 0.43–0.64 |  |  |  |

Table 4. ANC $A_{NC}$ values for the $p^{14}N$ channel from Ref. 27 with indication of range of values, average value and dimensional AC

| Level $^{15}$O, MeV | $J^\pi$ | $nlj$ | ANC $A_{NC}$, fm$^{-1}$ Ref. 27 | ANC $A_{NC}$, fm$^{-1}$ Ref. 45 | ANC $A_{NC}$, fm$^{-1}$ Ref. 46 | Interval of ANC $A_{NC}$ values, fm$^{-1}$ | Average on interval ANC $A_{NC}$, fm$^{-1}$ | Interval of the AC values, $C^2$, fm$^{-1}$ | Average on the interval AC $C^2$, fm$^{-1}$ | Average on the interval AC $C$, fm$^{-1/2}$ |
|---|---|---|---|---|---|---|---|---|---|---|
| 0 | 1/2$^-$ | $1p$ | 58.9(6.8) | 57(13) | 49(5)$^a$ | 44–70 | 57(13) {7.5(8) fm$^{-1/2}$} | 30–80 | 55(25) | 7.4(1.5) |
| 5.18 | 1/2$^+$ | $2s_{1/2}$ | 0.10(3) | 0.100(36) | --- | 0.065–0.135 | 0.100(35) | 6.5–13.5 (for $S_f$ = 0.01) 3.25–6.75 (for $S_f$ = 0.02) | 10.0(3.5) (for $S_f$ = 0.01) 5.0(1.75) (for $S_f$ = 0.02) | 3.2(4) (for $S_f$ = 0.01) 2.2(4) (for $S_f$ = 0.02) |
| 5.24 | 5/2$^+$ | $1d_{5/2}$ | 0.12(3) | 0.108(30) | 0.10(1) | 0.08–0.14 | 0.11(3) | 0.7–4.7 | 2.7(2.0) | 1.6(6) |
| 6.17 | 3/2$^-$ | $1p_{1/2}$ | 0.34(7) | 0.46(10) | 0.450(55) | 0.27–0.56 | 0.42(15) | 1.7–14 | 7.8(6.2) | 2.8(9) |
| 6.79 | 3/2$^+$ | $2s_{1/2}$ | 18.0(4.5) | 19(5) | 21.7(4.5) [24.5(6.1)]$^b$ | 13.5–30.6 | 22(8.5) | 24–113 | 68.5(44.5) | 8.3(3.4) |
| 6.86 | 5/2$^+$ | $1d_{5/2}$ | 0.44(7) | 0.32(7) | 0.290(36) | 0.25–0.51 | 0.38(13) | 0.4–1.4 | 0.9(5) | 0.9(3) |

a – the value of $C^2(p_{1/2}) + C^2(p_{3/2})$ is given, and for each constants in Ref. 46 $C^2(p_{1/2})$ = 39.0(5) fm$^{-1}$ and $C^2(p_{3/2})$ = 4.5(5) fm$^{-1}$ were obtained. These values were recalculated in Ref. 27 for $C^2(dp)$ = 4.32 fm$^{-1}$.
b – the ANC was obtained in Ref. 46 based on data of Ref. 45 and recalculated in Ref. 27 for $C^2(dp)$ = 4.32 fm$^{-1}$.

In addition, results for ANC $A_{NC}$ of new work Ref. 49 are known, which we present in a separate Table 5. These results, however, within limits of errors, coincide with the results from Refs. 27,45,46, shown in the previous Table 4.

Table 5. The ANC values for the $p^{14}N$ channel obtained in Ref. 49.

| Level $^{15}$O, MeV | $J^\pi$ | Channel spin $S$ | ANC $A_{NC}$, fm$^{-1}$ Ref. 49 | ANC $A_{NC}$, fm$^{-1/2}$ Ref. 49 |
|---|---|---|---|---|
| 0 | 1/2$^-$ | 1/2 | 0.05 | 0.23 |
| 0 | 1/2$^-$ | 3/2 | 54.8 | 7.4 |
| 5.18 | 1/2$^+$ | 1/2 | 0.11 | 0.33 |
| 5.24 | 5/2$^+$ | 1/2 | 0.05 | 0.23 |
| 5.24 | 5/2$^+$ | 3/2 | 0.06 | 0.24 |
| 6.17 | 3/2$^-$ | 1/2 | 0.22 | 0.47 |
| 6.17 | 3/2$^-$ | 3/2 | 0.28 | 0.53 |
| 6.79 | 3/2$^+$ | 1/2 | 7.3(1.1) | 2.7(2) |
| 6.79 | 3/2$^+$ | 3/2 | 24.1(7) | 4.91(7) |
| 6.86 | 5/2$^+$ | 1/2 | 0.15 | 0.39 |
| 6.86 | 5/2$^+$ | 3/2 | 0.18 | 0.42 |

Now we give the interval of possible values of dimensionless AC, which is obtained on basis of results given in the previous tables for the ANC and the spectroscopic factors of $S_f$.

Table 6. Dimensionless AC $C_w$ for the BS in the $p^{14}N$ channel of $^{15}O$.

| No. | Level $^{15}O$, MeV | $J^\pi$ | Average on interval AC $C$, fm$^{-1/2}$ | $\sqrt{2k_0}$ | Average on interval dimensionless AC $C_w$ |
|---|---|---|---|---|---|
| GS | 0 | $1/2^-$ | 7.4(1.5) | 1.07 | 6.9(1.4) |
| 1$^{st}$ES | 5.18 | $1/2^+$ | 3.2(4) for $S_f$ = 0.01<br>2.2(4) for $S_f$ = 0.02 | 0.79 | 4.0(5) for $S_f$ = 0.01<br>2.8(5) for $S_f$ = 0.02 |
| 2$^{nd}$ES | 5.24 | $5/2^+$ | 1.6(6) | 0.78 | 2.0(8) |
| 3$^{rd}$ES | 6.17 | $3/2^-$ | 2.8(9) | 0.67 | 4.2(1.5) |
| 4$^{th}$ES | 6.79 | $3/2^+$ | 8.3(3.4) | 0.55 | 15.1(6.2) |
| 5$^{th}$ES | 6.86 | $5/2^+$ | 0.9(3) | 0.53 | 1.7(6) |

Furthermore, the 1$^{st}$ES of $^{15}O$ in the $p^{14}N$ channel with $J^\pi = 1/2^+$ can be matched as $^2S_{1/2}$ and $^4D_{1/2}$ waves. In particular, for $^2S_{1/2}$ potential with FS of the 1$^{st}$ES at excitation energy $E_x = 5.18$ MeV, one can obtain the parameters shown in Table 2 under No. 2. With this potential, a binding energy of -2.11410 MeV, which totally coincides with experimental value,[26] a charge radius of 2.48 fm, and a dimensionless AC of 1.9(1) over interval of 10–15 fm were obtained. The scattering phase shift for such potential, if we take into account generalized Levinson theorem, smoothly decreases to 357° in energy range to 1 MeV. In this case, there are two bound states – one FS and one AS. Note that the parameters of this potential were refined to correctly describe available data for the $S$-factor of the proton capture on $^{14}N$ to this ES in the energy region below the first resonance, i.e., at the lowest energies. Therefore, value of its asymptotic constant is somewhat below lower limit of the possible values shown in Table 6, and not in middle of this range, as was done for GS. However, it is well known that accuracy of determination of spectroscopic factors is currently not very high and, if it is refined, it may turn out that the values obtained above totally agree with each other.

In the second case, the 1$^{st}$ES can be matched the $^4D_{1/2}$ wave with the FS, for which parameters are also presented in Table 2, but under No. 3. With such a potential, a binding energy of -2.11410 MeV, with a FDM accuracy of 10$^{-5}$ MeV,[42] totally coinciding with the experimental value,[26] a charge radius of 2.51 fm and a dimensionless AC of 0.4(1) in the interval of 6–20 fm were obtained. The scattering phase shift for a given potential smoothly decreases to 357° in energy region up to 1 MeV. Note that the AC value of this level, found above (see Table 6), was obtained under an assumption that this state is the $^2S_{1/2}$ wave, and the potential of the $^4D_{1/2}$ wave can lead to other values of AC. In this case, we failed to find data for the corresponding ANC and $S_f$ in the $D$ wave for this BS. Parameters of such a potential were also refined to correctly describe available experimental data for $S$-factor of the proton capture on $^{14}N$ to the first ES at energies below the first resonance.

For the potential of the 2$^{nd}$ES $^{2+4}D_{5/2}$ of $^{15}O$ in the $p^{14}N$ channel with an FS at $E_x = 5.2$ MeV and $J^\pi = 5/2^+$, parameters shown in Table 2 under No. 4 were found. With this potential, a binding energy of -2.05620 MeV, which totally coincides with experimental value,[26] a charge radius of 2.64 fm, and a dimensionless asymptotic constant 1.1(1) in interval of 10–15 fm were obtained. The scattering phase shift for such potential smoothly decreases to 359° in energy region up to 1 MeV. The parameters of the 2$^{nd}$ES potential were also refined to correctly describe the magnitude of the experimental S-factor at the lowest energies of 150–200 keV. Therefore, dimensionless AC of this potential is at lower boundary of possible value for this level, for which value 1.2 is given in Table 6.

Since, the 5$^{th}$ES $J^\pi = 5/2^+$ have the same momentum, we consider it next. The parameters given in Table 2 under No. 9 were found for potential of this $^{2+4}D_{5/2}$ state of $^{15}$O with FS at $E_x = 6.86$ MeV. The parameters of this potential were also refined to correctly describe magnitude of the experimental S-factor at the lowest energies of 300–500 keV. With such a potential, binding energy of -0.43770 MeV, which totally coincides with the experimental value,[26] the charge radius of 2.67 fm, and dimensionless asymptotic constant of 1.0(1) over the interval 10–15 fm were obtained. The obtained AC is at lower limit of possible values of the AC listed in Table 6. The nonresonance scattering phase shift for such a potential smoothly decreases to 359° in the energy range up to 1 MeV.

Furthermore, for potential of the bound 3$^{rd}$ES $^{2+4}P_{3/2}$ of $^{15}$O in the $p^{14}$N channel without FS at $E_x = 6.17$ MeV and $J^\pi = 3/2^-$, parameters are found, which are listed in Table 2 under No. 5. With this potential, a binding energy of -1.12080 MeV, which totally coincides with the experimental value,[26] a charge radius of 2.55 fm, and a dimensionless asymptotic constant 1.6(1) in interval of 10–15 fm were obtained. The scattering phase shift for such potential smoothly decreases to 178° in energy region up to 1 MeV. From the obtained results it can be seen that the AC of the potential of 3$^{rd}$ES is slightly below the lower limit of interval of values for the AC equal to 2.7, as shown in Table 6. However, as already mentioned, the range of $S_f$ values may turn out to be somewhat larger than the one given above, so interval of the possible value of dimensionless AC can also be extended.

This state can also be attributed to the bound $^4F_{3/2}$ wave without FS – for such an option of potential, Table 2 shows the parameters under No. 6. This potential leads to correct binding energy of the $p^{14}$N channel, a somewhat larger charge radius of 3.22 fm and a dimensionless AC equal to 2.3(1). We did not manage to find the AC values for this potential option, therefore it is difficult to judge the obtained value. The elastic scattering phase shift of this potential is in the region of 180.1(1) for energies up to 1 MeV.

For the first $^4S_{3/2}$ potential of the 4$^{th}$ES of $^{15}$O in the $p^{14}$N channel at $E_x = 6.7931$ MeV and $J^\pi = 3/2^+$, the potential parameters with FS, given in Table 2 under No. 7, were obtained. With such potential, a binding energy of –0.50400 MeV, totally coinciding with experimental value,[26] a charge radius of 2.86 fm and a dimensionless asymptotic constant of 8.8(1) over the interval of 10–20 fm, were obtained. This value agrees totally with the data of Table 6, and potential scattering phase shift smoothly drops to 350º at energy of 1 MeV.

For the 4$^{th}$ES, we can use another option, namely, $^{2+4}D_{3/2}$ wave, which allows one to obtain parameters of the potential with FS, which are given in Table 2 under No. 8. Such potential leads to a binding energy of –0.50400 MeV, which coincides with the experimental value,[26] a charge radius of 3.48 fm, and AC 7.1(1) on an interval of 8–20 fm. The obtained value of the asymptotic constant is somewhat below the lower limit of possible values given above in Table 6, which, however, were determined, as in the case of the 1$^{st}$ES, for the S wave. The scattering phase shift of this potential smoothly decreases to 358 ° at proton energy of 1 MeV.

## 4. Astrophysical *S*-factors for the proton capture on $^{14}$N to the BS of $^{15}$O

Let us now consider results of calculations for the astrophysical *S*-factor in case of the proton capture on $^{14}$N to five excited states and the ground state of $^{15}$O. First, we will focus on the capture to the 4$^{th}$ES, since, as already mentioned, such capture makes the largest contribution to the total *S*-factor of the $^{14}$N$(p,\gamma)^{15}$O capture process.

### 4.1. *Astrophysical S-factor for capture to the 4$^{th}$ES*

Let us consider here possible transitions between 4$^{th}$ES and different scattering states. Previously we considered it in Ref. 4. Also we give a method for calculating total cross sections for states mixed by the spin of channel *S*.

#### 4.1.1. *Transitions for the proton capture on $^{14}$N to the 4$^{th}$ES*

If we match the quartet $^4S_{3/2}$ wave to the fourth bound ES, then we can consider *E*1 transitions from all quartet $^4P$ scattering waves, which do not have FS. As a first option of the potentials of such *P* waves (the second is discussed later), it is assumed that the potential has zero depth, as shown in Table 1, leading to zero scattering phase shifts, then these transitions are possible

$$^4P_{1/2} \to {}^4S_{3/2}; \quad {}^4P_{3/2} \to {}^4S_{3/2}; \quad {}^4P_{5/2} \to {}^4S_{3/2}.$$

The astrophysical *S*-factor for such transitions is smooth and plays role of non-resonance capture, ensuring correct behavior of the *S*-factor at the lowest energies. However, if we assume that resonance at 260 keV is the $^2S_{1/2}$ scattering wave, the first peak of *S*-factor cannot be explained – transition from such a wave is impossible along the spin of the channel. If for this resonance with $J = 1/2^+$ the $^4D_{1/2}$ wave is used and for $J = 3/2^+$ the $^4D_{3/2}$ scattering wave, *E*2 transitions to the 4$^{th}$ES are possible.

$$\begin{aligned}^4D_{1/2} &\to {}^4S_{3/2} \\ {}^4D_{3/2} &\to {}^4S_{3/2}\end{aligned}.$$

However, the magnitude of the first resonance, as shown below, is small, and it cannot explain the available experimental data. Thus, if we compare quartet $^4S_{3/2}$ wave to the 4$^{th}$ES, it is not possible to describe magnitude of the main resonance at 260 keV at all.

At the same time, if we consider the transition to the fourth excited state with $J^\pi = 3/2^+$, by matching it with a mixture of $^{2+4}D_{3/2}$ waves, then we can consider *M*1 process

$$^4D_{1/2} \to {}^4D_{3/2},$$

which leads to large cross sections at 260 keV. In addition, *E*1 processes from *P* scattering waves are also possible.

$$^2P_{1/2} \to {}^2D_{3/2};\ {}^2P_{3/2} \to {}^2D_{3/2};$$
$$^4P_{1/2} \to {}^4D_{3/2};\ {}^4P_{3/2} \to {}^4D_{3/2};\ {}^4P_{5/2} \to {}^4D_{3/2}$$

which, as in past case, perform role of nonresonance capture. In this case, transitions occur between $^{2+4}P$ mixed scattering states and also mixed by spin $^{2+4}D_{3/2}$ 4$^{th}$ES. Therefore, total cross section is written in form of a simple sum, i.e.

$$\sigma(E1) = \sigma(^2P_{1/2} \to {}^2D_{3/2}) + \sigma(^4P_{1/2} \to {}^4D_{3/2})$$

or

$$\sigma(E1) = \sigma(^2P_{3/2} \to {}^2D_{3/2}) + \sigma(^4P_{3/2} \to {}^4D_{3/2})$$

can be considered as a simple doubling of the cross section.

Since the model used does not allow one to explicitly separate spin states, so the same wave functions mixed by spin and found in a given potential are used to calculate each part of such a process. As a result, only spin factors are different in such matrix elements. In reality, there is only one transition from scattering states to the bound 4$^{th}$ES of the nucleus, but not two different $E1$ transitions. Therefore, total cross section must be represented in form of averaging over transitions from the spin mixed $P$ scattering waves to the mixed $D_{3/2}$ 4$^{th}$ES of $^{15}$O in the $p^{14}$N channel. Such an approach, which we proposed earlier when considering transitions in the neutron radiative capture on $^{14}$N,[2] was also used for some other reactions and allowed us to obtain good results in describing total cross sections of the capture reactions.[1,2,22]

Therefore, total cross section of the capture process to the 4$^{th}$ES for electromagnetic $E1$ transitions is represented as the following combination of partial total cross sections

$$\sigma(E1) = \sigma(^4P_{5/2} \to {}^4D_{3/2}) + [\sigma(^2P_{1/2} \to {}^2D_{3/2}) + \sigma(^4P_{1/2} \to {}^4D_{3/2})]/2 +$$
$$+ [\sigma(^2P_{3/2} \to {}^2D_{3/2}) + \sigma(^4P_{3/2} \to {}^4D_{3/2})]/2$$

Here, the averaging over transitions with the same total momentum, but different channel spin is performed. For M1 transitions, you also need to perform averaging over spin of channel.

$$\sigma(M1) = \sigma(^4D_{1/2} \to {}^4D_{3/2}) + [\sigma(^2D_{5/2} \to {}^2D_{3/2}) + \sigma(^4D_{5/2} \to {}^4D_{3/2})]/2 +$$
$$+ [\sigma(^2D_{3/2} \to {}^2D_{3/2}) + \sigma(^4D_{3/2} \to {}^4D_{3/2})]/2$$

Furthermore, in Table 7,[4] $P^2$ coefficients are given in above expressions for the total cross sections and for all the transitions considered below and for two options of partial waves for the 4$^{th}$ES.

Table 7. $P^2$ coefficients in total cross sections for considered transitions to the 4$^{th}$ES.

| No. | $[^{(2S+1)}L_J]_i$ | Type of transition | $[^{(2S+1)}L_J]_f$ | $P^2$ |
|---|---|---|---|---|
| 1 | $^4P_{1/2}$ | E1 | $^4S_{3/2}$ | 2 |
| 2 | $^4P_{3/2}$ | E1 | $^4S_{3/2}$ | 4 |
| 3 | $^4P_{5/2}$ | E1 | $^4S_{3/2}$ | 6 |
| 4 | $^4D_{1/2}$ | E2 | $^4S_{3/2}$ | 2 |
| 5 | $^4D_{3/2}$ | E2 | $^4S_{3/2}$ | 4 |
| 6 | $^2P_{1/2}$ | E1 | $^2D_{3/2}$ | 4 |
| 7 | $^2P_{3/2}$ | E1 | $^2D_{3/2}$ | 4/5 |
| 8 | $^4P_{1/2}$ | E1 | $^4D_{3/2}$ | 2 |
| 9 | $^4P_{3/2}$ | E1 | $^4D_{3/2}$ | 64/25 |
| 10 | $^4P_{5/2}$ | E1 | $^4D_{3/2}$ | 6/25 |
| 11 | $^4D_{1/2}$ | M1 | $^4D_{3/2}$ | 6 |
| 12 | $^4D_{1/2}$ | E2 | $^4D_{3/2}$ | 2 |
| 13 | $^4D_{5/2}$ | M1 | $^4D_{3/2}$ | 42/5 |
| 14 | $^2D_{5/2}$ | M1 | $^2D_{3/2}$ | 12/5 |
| 15 | $^4D_{3/2}$ | M1 | $^4D_{3/2}$ | 3/5 |
| 16 | $^2D_{3/2}$ | M1 | $^2D_{3/2}$ | 3/5 |

Now note that if we use the same potentials for the $^{2+4}D_{3/2}$ scattering wave and for the bound 4$^{th}$ES, for example, No. 8 from Table 2, then the matrix element for the M1 transitions of the form $^{2+4}D_{3/2} \to {}^{2+4}D_{3/2}$ are zero due to orthogonality of wave functions of these states. Similar M1 transition like $^4S_{3/2} \to {}^4S_{3/2}$ is also equal to zero, if we use for both of these states, for example, potential No. 7 from Table 2. It is possible that this is why in experimental measurements of total cross sections for the capture at 4$^{th}$ES, and then, in the astrophysical S-factor, such resonance is not observed.

However, it should be noted here that the potential of the bound $^{2+4}D_{3/2}$ state No. 8 from Table 2 is clearly different from the $^{2+4}D_{3/2}$ scattering potentials No. 4–No. 6 from Table 1. Each of this scattering potentials were constructed to describe its characteristics, for example, resonance parameters, and each BS potential were determined on basis of reproducing properties of the given state in discrete spectrum. For such "identical" states that have the same JLS, it is impossible to build the same potentials for describing scattering processes and BS. At least, this is true for Gaussian type of potentials. Therefore, we will further consider a version of calculations in which different potentials are used for such states of continuous and discrete spectrum.

### 4.1.2. *Astrophysical S-factor for capture to the 4$^{th}$ES*

First of all, an attempt was made to describe cross sections and S-factor of the radiative capture to the 4$^{th}$ES using processes No. 1–No. 5 from Table 7, i.e. when using for it $^4S_{3/2}$ waves with potential No. 7 from Table 2. The E1 transitions from all $^4P$ scattering waves with zero depth potentials were considered – results are presented in Fig. 2a with a red dashed curve. Furthermore, E2 transition from the resonance $^4D_{1/2}$ scattering wave No. 4 from Table 7 was considered. Parameters No. 2 from Table 1 were used for resonance scattering potential, which allow to correctly reproduce position and width of the first resonance at 260 keV. The results of such calculation for resonance in the $^4D_{1/2}$ wave are shown in Fig. 2a by the green dotted curve, and total $S$-factor, taking into account transitions from the $P$ waves, is shown by the blue solid curve. These results almost correctly describe experimental data at energies up to 3–3.5 MeV, excluding resonance region. Transition No. 5 from Table 7 with potential No. 4 from Table 1 does not make any real contribution to values of the total $S$-factor, as observed in the experiment.

As it can be seen from Fig. 2a, the calculated cross sections have a resonance, but its value is substantially lower than the results of experimental measurements from Refs. 50–54 presented in Fig. 2a with different points. $S$-factor in the region of 50–150 keV is equal to 1.33(8) keV·b, and at 30 keV it has a value of 1.50(1) keV·b. At energies below 50 keV, calculated $S$-factor experiences a rise due to a more rapid growth of permeability factor. Therefore, energies of 50–150 keV·b can be considered as an area of stabilization of its values, which can be used to linearly extrapolate results to zero energy. In fact, its value in stabilization region can be considered as an $S$-factor at zero energy, i.e., 1.33(8) keV·b.

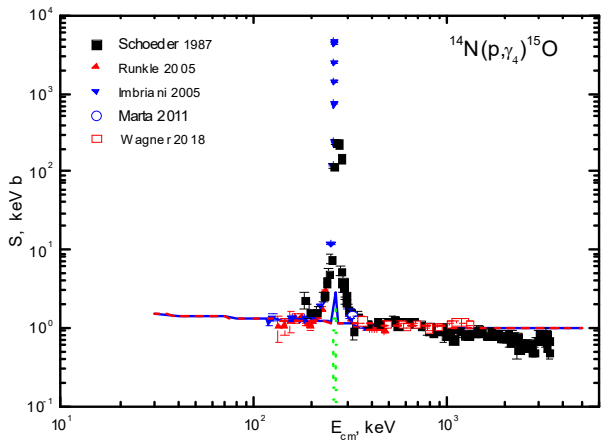 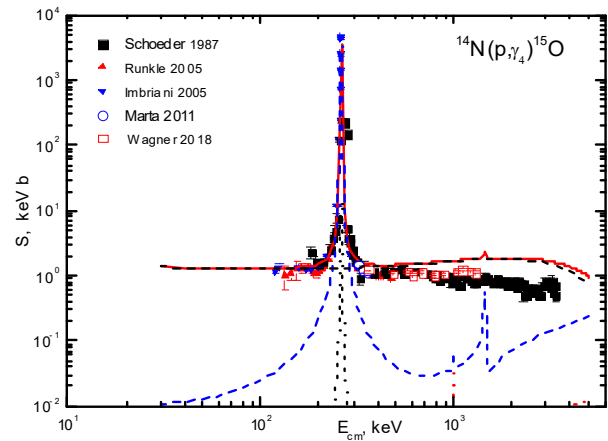

Fig.2a. Total cross section of the proton radiative capture on $^{14}$N to the 4$^{th}$ES at excitation energy of 6.79 MeV of $^{15}$O. Experimental data are taken from Refs. 50–54. Curves – results of our calculations, described in the text.

Fig.2b. Total cross section of the proton radiative capture on $^{14}$N to the 4$^{th}$ES at excitation energy of 6.79 MeV of $^{15}$O. Experimental data are taken from Refs. 50–54. Curves – results of our calculations, described in the text.

At resonance, $S$-factor reaches approximately 3 keV·b, which is three orders of magnitude less than the results of experimental measurements[50] and about two orders of magnitude less than data of Ref. 51. Experimental value of the $S$-factor from Ref. 50 is 4500 keV·b, and according to Ref. 51 this maximum reaches only 230–240 keV·b. Thus, it can be seen that $E2$ transition of type No. 4 from Table 7 is not capable of correctly describing magnitude of the $S$-factor at resonance energy. At the same time, transitions No. 1,2,3 from Table 7 make it possible to correctly give its value in nonresonance energy range up to 3–3.5 MeV.

Furthermore, if we use the $^{2+4}D_{3/2}$ potential for the bound 4$^{th}$ES No. 8 from Table 2, then for the total $S$-factor we get results shown in Fig. 2b by the red solid curve – here we considered transitions No. 6 to No. 14 from Table. 7 The black dashed curve shows results for $E1$ transitions from $P$ scattering waves No. 6–10 from Table 7 with potential No. 9 from Table 1. For the resonance $^4D_{1/2}$ scattering potential in transitions No. 11,12 from Table 7, parameters No. 2 from Table 1 are still used. For the second resonance in this wave, potentials No. 3 from Table 1 were used, and for nonresonance

$^{2+4}D_{5/2}$ potentials, at transitions No. 13,14 from Table 7, parameters No. 8 from Table 1. All these results are shown in Fig. 2b by the blue dashed curve. Processes No. 13,14 with potentials No. 8 of Table 1 lead to a smoothly increasing $S$-factor with a magnitude of about 0.2–0.3 keV·b at an energy of 5 MeV. Here, transition from the second resonance in the $^4D_{1/2}$ wave with potential No. 3 in Table 1 was taken into account. Accounting for this resonance leads to results that give a small maximum in resonance region at 1447 keV. The $E2$ process No. 12 from Table 7, considered here, makes a contribution only at resonance energy and has a value less than 15 keV·b, at all other energies its contribution is negligible. The $S$-factor of this process is shown in Fig. 2b by the black dotted curve.

Calculations in Fig. 2b make it possible to correctly reproduce cross sections at the lowest energies up to 150–200 keV and correctly describe values of near-resonance $S$-factor in the energy range of 150–250 keV and 270–320 keV. At resonance, calculated $S$-factor reaches a value of about 3700 keV·b, in region of 50–150 keV it is equal to 1.32(2) keV·b, and at 30 keV it has a value of 1.37 keV·b. However, at energies greater than resonance, the total cross sections go somewhat higher than available data.

Furthermore, for comparison, the astrophysical $S$-factor at zero energy, with results of other works, reproduce Table 8 from Ref. 27 with the addition of new results from Refs. 49,54. This table shows that our $S(0)$-factor value of 1.32(2) keV·b agrees well with results of other works, which give an average value of 1.31 keV·b and a range from about 1.1 to 1.8 keV·b with an average over range of 1.45(35) keV·b.

Table 8. Comparison of results for $S$-factors at zero energy for the proton radiative capture on $^{14}$N to the 4$^{th}$ES

| Refs. | 27 | 50 | 52 | 46 | 31 | 49 | 54 | Average |
|---|---|---|---|---|---|---|---|---|
| $S(0)$, keV·b | 1.26(17) | 1.20(5) | 1.15(5) | 1.4(2) | 1.63(17) | 1.29(15) | 1.24(13) | 1.31 |

This Table was published in our work Ref. 4.

Thus, from results above it can be seen that it is possible to find intercluster potential of the 4$^{th}$ES (No. 8 of Table 2), if we take it as $^{2+4}D_{3/2}$ wave, which as a whole, allows us to correctly describe available experimental data for the astrophysical $S$-factor of the proton radiative capture on $^{14}$N to the 4$^{th}$ES of $^{15}$O at energy $E_x = 6.79$ MeV. At the same time, parameters No. 2 from Table 1 were used for resonance scattering potential, which make it possible to correctly reproduce position and width of the first resonance at 260 keV.

An increase in $S$-factor values at energies above resonance can be explained, for example, by following. We were unable to find the results of phase shift analysis for the elastic $p^{14}$N scattering. Therefore, our potentials for $P$ waves cannot be constructed quite correctly. At the very beginning of the article it was assumed that since the spectra of $^{15}$O do not have negative parity resonances, and $P$ waves do not have FS, then they can be matched to the potentials of zero depth, leading to zero scattering phase shifts. However, when performing phase shift analysis, it may turn out that these phase shifts have a smoothly decreasing or increasing character at energies greater than energy of the first resonance, which can noticeably change the behavior of the calculated $S$-factor in this energy range. Performing such an analysis can contribute to a more correct construction of potentials of the $p^{14}$N interactions in the nonresonance energy range and it will allow more correctly carry out similar calculations of cross sections at energies above the first resonance.

Despite the fact that transitions of type's No. 15,16 in Table 7 should be forbidden for identical potentials, as mentioned above, we still made calculation of such an $M1$ transition with averaging over the spin of the channel. In this case, different potentials are used for the scattering processes and BS, which were constructed on basis of description of characteristics of these states. The results of such a calculation are presented at the very bottom of Fig. 2b by the red dotted curve. At resonance 987 keV, value of this $S$-factor reaches 0.03 keV·b and practically does not affect the total $S$-factor, which at this energy has a value of order 1.5–2.0 keV·b.

Thus, despite use of different potentials of continuous and discrete spectra, we obtain effective, numerical orthogonality of these states with an accuracy of order of 2%. In other words, potentials based on description of observed characteristics of continuous and discrete spectra turn out to be constructed in such a way that they naturally lead to orthogonality of the similar states, i.e., states with same $JLS$ but different signs of energy of these states.

Now consider possibility of using certain BS potentials for scattering potentials. For $P_{1/2,3/2}$ waves, potentials were obtained for the GS No. 1 and for the 3$^{rd}$ES No. 5 from Table 2, which correctly describe characteristics of these states. On basis of general principles we can use such potentials to describe scattering processes in these partial waves and only for the $P_{5/2}$ wave, since there are no observable bound states in this wave, leave the parameters No. 9 from Table 1. There are resonances in the $D_{1/2}$ and $D_{3/2}$ scattering waves, and it is not possible to combine potential of continuous and discrete spectrum. Instead of the $D_{5/2}$ scattering No. 8 from Table 1, one can use potential, for example, 2$^{nd}$ES No. 4 from Table 2. As a result, if these parameters are used for $D_{5/2}$ and $P$ waves, neither general picture of the $S$-factor description, nor its values at the lowest energies given above, do not change. Thus, presence in additional $P$ waves and $D$ wave of an additional allowed BS does not lead to any noticeable changes in results of calculation of the $S$-factor during radiative capture to the 4$^{th}$ES.

### 4.2. *Astrophysical S-factor for capture to the second and fifth ES*

We now consider results of calculations for the astrophysical $S$-factor in case of the proton capture on $^{14}$N to second and fifth excited states of $^{15}$O. Previously we considered them also in Ref. 8. These two states are considered together, since they have the same moments and parity of $J^{\pi} = 5/2^{+}$ and are most likely related to the spin-mixed partial $^{2+4}D_{5/2}$ wave.

### 4.2.1. *Transitions at the capture to the second and five ES*

The second and fifth ES have a moment of 5/2$^{+}$ and can relate to the $^{2+4}D_{5/2}$ wave, therefore, $E1$ transitions are possible here

$$^{2}P_{3/2} \to {}^{2}D_{5/2}$$
$$^{4}P_{3/2} \to {}^{4}D_{5/2},$$
$$^{4}P_{5/2} \to {}^{4}D_{5/2}$$

with averaging, which was mentioned in previous paragraph for other states

$$\sigma(E1) = \sigma(^4P_{5/2} \to {}^4D_{5/2}) + [\sigma(^2P_{3/2} \to {}^2D_{5/2}) + \sigma(^4P_{3/2} \to {}^4D_{5/2})]/2,$$

as well as $M1$ processes from the resonance $^{2+4}D_{3/2}$ wave when averaged over spin of the channel

$$\sigma(M1) = \sigma(^4D_{7/2} \to {}^4D_{5/2}) + [\sigma(^2D_{3/2} \to {}^2D_{5/2}) + \sigma(^4D_{3/2} \to {}^4D_{5/2})]/2$$

Here, averaging over transitions with the same total moment, but different channel spin is also carried out.

As in the previous case (see Table 7), for transitions from $^{2+4}D_{5/2}$ scattering waves, matrix elements are considered to be equal to zero due to orthogonality of the WF of discrete and continuous spectrum when using the same potentials for them, for example, 2$^{nd}$ES or 5$^{th}$ES. The partial $^{2+4}D_{5/2}$ scattering wave has no resonances, we consider its phase shift close to zero, and not only parameters No. 8 from Table 1 can be used for it, but also parameters No. 4 and No. 9 from Table 2 for BS potentials.

Table 9. $P^2$ coefficients in total cross sections for transitions to the second and fifth ES

| No. | $[^{(2S+1)}L_J]_i$ | Type of transition | $[^{(2S+1)}L_J]_f$ | $P^2$ |
|---|---|---|---|---|
| 1 | $^2P_{3/2}$ | $E1$ | $^2D_{5/2}$ | 36/5 |
| 2 | $^4P_{3/2}$ | $E1$ | $^4D_{5/2}$ | 126/25 |
| 3 | $^4P_{5/2}$ | $E1$ | $^4D_{5/2}$ | 54/25 |
| 4 | $^2D_{3/2}$ | $M1$ | $^2D_{5/2}$ | 12/5 |
| 5 | $^4D_{3/2}$ | $M1$ | $^4D_{5/2}$ | 42/5 |
| 6 | $^4D_{7/2}$ | $M1$ | $^4D_{5/2}$ | 48/7 |
| 7 | $^4D_{1/2}$ | $E2$ | $^4D_{5/2}$ | 6/7 |
| 8 | $^2D_{5/2}$ | $M1$ | $^2D_{5/2}$ | 21/10 |
| 9 | $^4D_{5/2}$ | $M1$ | $^4D_{5/2}$ | 507/70 |

Above, the summary Table 9 (it is earlier given in our work Ref. 8) of all $E1$, $M1$ and one $E2$ transitions from resonance $^4D_{1/2}$ wave to the 2$^{nd}$ES or the 5$^{th}$ES and coefficients $P^2$ in expressions for total cross sections are given.

For $P_{3/2,5/2}$ waves of continuous spectrum, parameters of potential, as the first option, are considered to be equal to zero, as given in Table 1 under No. 9. As a second option of the parameters for the $^{2+4}P_{3/2}$ scattering waves, potential of the 3$^{rd}$ES No. 5 from Table 2 are used.

### 4.2.2. *Astrophysical S-factor for capture to the 2$^{nd}$ES*

First, with potentials No. 2–No. 6 shown in Table 1, which have resonances in $D$-scattering waves, the $S$-factor was calculated for the proton radioactive capture on $^{14}$N at energies up to 5 MeV. In the first version of calculations, potential No. 8 from Table 1 was used $^{2+4}D_{5/2}$ waves. Recall that total cross sections for transitions from $^{2+4}D_{5/2}$ scattering waves to the 2$^{nd}$ES are considered to be equal to zero, since these states must be orthogonal if the same potentials are used for them. But in this case potentials of continuous and discrete spectra are different and it is necessary to show in numerical form that such transitions will not make a significant contribution to the $S$-factor of the radiative capture reaction to the 2$^{nd}$ES.

Furthermore, in Fig. 3a the capture results from *P* scattering waves of No. 1-3 from Table 9 with potentials of zero depth No. 9 from Table 1 are shown by the green dotted curve. The blue dashed curves show the results for the *M*1 and *E*2 captures from all *D* waves (No. 4–9 in Table 9), red dotted curve shows the result for capture from $^4D_{7/2}$ waves, blue dotted curve shows transitions from the $^{2+4}D_{5/2}$ waves, and the red solid curve gives total result for astrophysical *S*-factor of the proton capture process on $^{14}$N to the second excited state of $^{15}$O, which is bound in the $p^{14}$N channel.

As it can be seen from calculated *S*-factor shown in Fig. 3a, its value has a resonance at energy of 260 keV (*E*2 transition No. 7 from Table 9) and 987 keV (transitions No. 4,5 from Table 9) and, in general, experimental data are correctly described, which are taken from Refs. 50,51. The first maximum is somewhat less than data of Ref. 50, and the value of the second is slightly larger than measurement results from Ref. 51. The calculated minimum between resonances is in good agreement with experimental results[51] and is totally determined by *S*-factor for capture to the second ES from the *P* scattering waves No. 1–3 from Table 9.

For transitions from other $D_{3/2}$ scattering waves with potentials No. 5,6 from Table 1, resonances are also observed and heights of both are noticeably greater than the measurements from Ref. 51. The *S*-factors of the *E*2 transitions from $^4D_{1/2}$ waves at 260 and 1447 keV are shown in Fig. 3a with the green dashed curve. The calculated resonance for the *E*2 process at 1447 keV with potential No. 3 from Table 1, which was not observed in the experiment,[51] has almost no significant contribution to the total *S*-factor. The contributions from $^{2+4}D_{5/2}$ wave transitions are 0.015 keV·b at 5 MeV and do not lead to any changes in total the *S*-factor of this process.

In other words, this process leads to numerical results for the *S*-factor tending to zero. Here, apparently, it should be repeated that potentials based on description of observed characteristics of scattering and BS processes are constructed in such a way that they usually lead to numerical orthogonality of identical states, i.e., states with same *JLS* in continuous and discrete spectrum. In other words, contribution of such a transition is order of 1–2% at the highest energies of 5 MeV and does not lead to a real change in magnitude or shape of the *S*-factor.

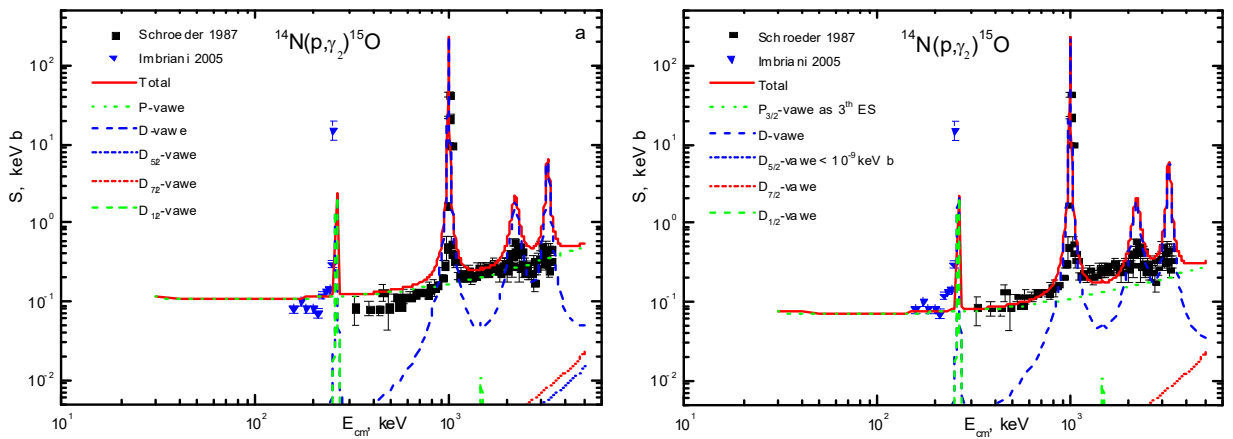

Fig.3a,b. Astrophysical *S*-factor of the proton radiative capture on $^{14}$N to the 2$^{nd}$ES at 5.24 MeV of $^{15}$O. The experimental data are taken from Refs. 50,51. Curves – results of our calculations which are described in the text.

The calculated *S*-factor shown in Fig. 3a has a value of 0.11(1) keV·b at energy of 50–200 keV, and 0.11 keV·b at 30 keV. The error is determined by the averaging of the *S*-factor in specified energy range. For comparison, we present some experimental values of the *S*-factor at zero energy for capture to the second ES. In Ref. 27, value 0.066(25) keV·b was obtained, in Ref. 50 0.070(3) keV·b was found for the *S*(0)-factor. This shows that value of the calculated *S*-factor obtained above is not very different from results of these works.

As a second version of calculations, we use instead of the $D_{5/2}$ scattering potential with parameters No. 8 from Table 1, potential of the $2^{nd}$ES No. 4 from Table 2. In this case, calculated *S*-factor for this transition to the $2^{nd}$ES turns out to be less than $10^{-9}$ keV·b in entire energy range. This result well demonstrates numerical orthogonality of continuous and discrete spectrum states in one potential. In energy region of our interest, on average, i.e., without resonances, calculated *S*-factor has a value of 0.1–0.5 keV·b, therefore, for indicated transition, its relative value is at level of $10^{-8}$.

Furthermore, if real potential of the $3^{rd}$ES No. 5 from Table 2 is used for the $P_{3/2}$ scattering wave (potential No. 9 from Table 1 is still used for the $P_{5/2}$ wave, since there is no bound allowed state for it), then the general form of description of experimental data remains almost unchanged. However, results for the *S*-factor with capture from *P* scattering waves, shown in Fig. 3b by the green dotted curve, are somewhat reduced.

For *S*-factor at 30 keV, 0.074 keV·b is obtained, in energy interval of 50–200 keV its values are equal to 0.72(2) keV·b – this value can be considered as its value is at zero energy. Such results are in better agreement with data of Ref. 50, where 0.070(3) keV·b is presented. Thus, using correct potential of the $P_{3/2}$ scattering wave, results are obtained that are in better agreement with available experimental data.

### 4.2.3. *Astrophysical S-factor of the capture to the $5^{th}$ES*

Let us now consider *E*1, *E*2, and *M*1 transitions to fifth ES of $^{15}$O at the radiative capture on $^{14}$N. The experimental data at all energies were taken from Ref. 51 and, for comparison, results for the *S*-factor at zero energy are given from Refs. 27,46. For potential of the bound fifth excited $^{2+4}D_{5/2}$ state of $^{15}$O in the $p^{14}$N channel with FS at $E_x$ = 6.86 MeV, parameters No. 9 from Table 2 are given. The types of transitions are shown in Table 9, and results of calculation the total capture *S*-factor are shown in Fig. 4a by the red solid curve. The blue dashed curve shows calculation results for the capture from all resonance *D* waves, and green dotted curve shows the results for the capture from *P* scattering waves with zero potential No. 9 from Table 1.

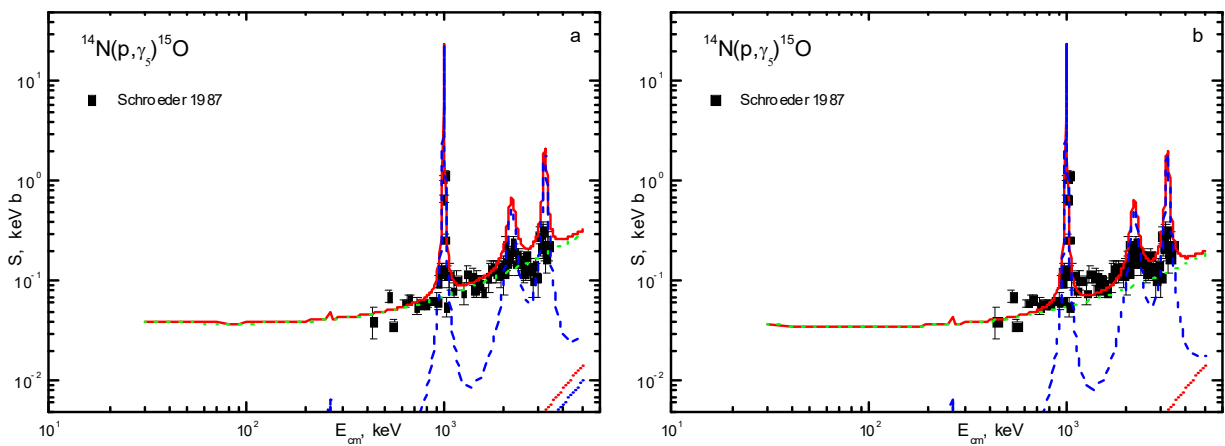

Fig.4a,b. Astrophysical *S*-factor of the radiative capture on $^{14}$N to the fifth ES at 6.86 MeV of $^{15}$O. Experimental data are taken from Ref. 51. Curves are results of our calculations; notations, as in Fig. 3a.

In all these calculations, same scattering potentials were used as in previous case for the first option of calculations in capture to the $2^{nd}$ES. The contribution of transitions from the $^{2+4}D_{5/2}$ waves is shown in the right lower corner of Fig. 4a by the blue dotted curve and has a value of 0.01 keV·b at 5 MeV, and the red dotted curve shows result for the capture from $^{4}D_{7/2}$ wave. For transitions from other $D$ waves of scattering with potentials No. 5,6 from Table 1, resonances are also observed, and the heights of both are noticeably larger than available measurements from Ref. 51. The first resonance at 260 keV, which is obtained due to $E2$ transition No. 7 from Table 9, is now hardly noticeable and, in fact, is within error band of the first experimental data point of Ref. 51. Its value is only 0.007 keV·b and strongly depends on level of binding energy, which in this case is about five times less than the energy of the $2^{nd}$ES, as given in Table 2.

Calculated $S$-factor shown in Fig. 4a at an energy range of 50–200 keV has a value of 0.039(1) keV·b, which can apparently be considered as its value at zero energy, and at 30 keV it is equal to 0.04 keV·b. The error is determined by the averaging of $S$-factor in the specified energy region. In Ref. 27, the $S$-factor for zero energy is given when it is captured to the fifth ES and 0.042(7) keV·b is obtained, and in Ref. 46 0.03(4) keV·b is found. This shows that value obtained by us agrees well with the results of Ref. 27 and is within limits of the errors of Ref. 46.

Thus, for this reaction, on basis of the approach used, it is quite possible to correctly describe behavior of the experimental $S$-factor for transitions to the $2^{nd}$ES and $5^{th}$ES of $^{15}$O in the $p^{14}$N channel.[8] The equality to zero of the depth of $P$ potentials leads to the correct description of the $S$-factor between resonances and above the second resonance, practically not overestimating its value, as it was in the case of capture to the $4^{th}$ES.

One has to use the potentials of $D$ scattering waves for two resonance scattering states, as in previous cases, which differs from results of, for example, Ref. 27 and comparisons given in it with other works. It was previously believed that resonance $S$ scattering waves correspond to these two states.[27,47,48,50,51] But as already mentioned, for example, in the $S$ wave it is generally impossible to construct resonance potential at 987 keV with required characteristics.

Here once again it is necessary to note that parameters of the potentials of the $2^{nd}$ES No. 4 and $5^{th}$ES No. 9 from Table 2 were refined for a more correct description of the $S$-factors at low energies. In essence, the task was to find such a BS potential, which is able, in general, to correctly describe all main characteristics of this state and $S$-factor at low energies. At the same time, parameters of potentials of resonance scattering waves were constructed solely to correctly describe characteristics of such resonances. In a process of calculating the astrophysical $S$-factor of the proton capture on $^{14}$N, their parameters did not change at all.

Furthermore, if real potential of the $3^{rd}$ES No. 5 from Table 2 is used for the $^{2+4}P_{3/2}$ waves, then overall picture of the $S$-factor description does not change, only its value at 30 keV becomes equal to 0.037 keV·b, and in energy range 50–200 keV varies in interval 0.035(1) keV·b. Using instead of the $D_{5/2}$ scattering wave No. 8 from Table 1, potential of the fifth ES No. 9 from Table 2 leads to the $S$-factor of this transition of the order of $10^{-9}$ keV·b in entire energy range. Here we see numerical orthogonality of such states. These results are shown in Fig. 4b with same curves as in Fig. 4a. In this case, it can be seen that use of real $P_{3/2}$ potential No. 5 from Table 2 as scattering potential for this wave does not lead to noticeable changes in shape and magnitude of the *S*-factor in considered energy range.

### 4.3. *Astrophysical S-factor for capture to the first ES*

Let us now consider the radiative proton capture on $^{14}$N to the first excited state of $^{15}$O. Previously we considered it in Ref. 5. Such state has total moment $J = 1/2^+$ and can relate to either $^2S_{1/2}$ or $^4D_{1/2}$ wave, therefore both options are considered later.

#### 4.3.1. *Capture transitions to the first ES*

In first version of calculations, we can consider transitions to the first excited $^2S_{1/2}$ state from doublet $^2P$ scattering waves whose potential is zero.

$$^2P_{1/2} \to\, ^2S_{1/2}$$
$$^2P_{3/2} \to\, ^2S_{1/2}$$

and *M*1 transition of the form

$$^2S_{1/2} \to\, ^2S_{1/2},$$

if we assume that the first resonance is exactly in the $^2S_{1/2}$ wave. We can assume that such states must be orthogonal when using the same potentials for continuous and discrete spectrum. However, in this case, the first resonance would simply not be observed, and experiment leads to existence of such a resonance. Therefore, in this wave there are states of continuous and discrete spectrum, and the question of their orthogonality is not so unique.

However, if we compare the first ES with a pure by spin $^4D_{1/2}$ wave, then *E*1 transitions from $^4P$ waves with zero potential are possible.

$$^4P_{1/2} \to\, ^4D_{1/2},$$
$$^4P_{3/2} \to\, ^4D_{1/2},$$

and also *M*1 processes from the *D* resonance waves

$$^4D_{1/2} \to\, ^4D_{1/2}.$$
$$^4D_{3/2} \to\, ^4D_{1/2}.$$

The question of orthogonality of the $^4D_{1/2}$ states of continuous and discrete spectrum are considered further. Since in both cases ES is pure by spin, averaging used earlier is absent here. Below is a summary Table 10 of all *E*1, *M*1 and some *E*2 transitions to the first ES and coefficients $P^2$ in expressions for total sections that were given above.

Table 10. $P^2$ coefficients for total cross sections at transitions to the first ES

| No. | $[^{(2S+1)}L_J]_i$ | Transition type | $[^{(2S+1)}L_J]_f$ | $P^2$ |
|---|---|---|---|---|
| 1 | $^2P_{1/2}$ | E1 | $^2S_{1/2}$ | 2 |
| 2 | $^2P_{3/2}$ | E1 | $^2S_{1/2}$ | 4 |
| 3 | $^2S_{1/2}$ | M1 | $^2S_{1/2}$ | 3/2 |
| 4 | $^2D_{3/2}$ | E2 | $^2S_{1/2}$ | 2 |
| 5 | $^2D_{5/2}$ | E2 | $^2S_{1/2}$ | 6 |
| 6 | $^4P_{1/2}$ | E1 | $^4D_{1/2}$ | 2 |
| 7 | $^4P_{3/2}$ | E1 | $^4D_{1/2}$ | 2/5 |
| 8 | $^4D_{1/2}$ | M1 | $^4D_{1/2}$ | 1 |
| 9 | $^4D_{3/2}$ | M1 | $^4D_{1/2}$ | 6 |

This Table was published in our work Ref. 5.

The potentials of these partial waves in scattering processes are listed in Table 1, and for BS in Table 2.

### 4.3.2. *Astrophysical S-factor of the capture to the 1st ES*

Furthermore, for capture to the first ES, we will consider two options for calculation of the astrophysical *S*-factor with different types of potentials of the first ES, given in Table 2.

*First option of calculations*

When considering the radiative capture, we first proceed from assumption that the resonance at 260 keV and the first ES are related to the $^2S_{1/2}$ wave. In accordance with the general principles, wave functions of the BS and scattering process for the same $^2S_{1/2}$ state in the same potential must be orthogonal and transition of the *M*1 type No. 3 from Table 10 cannot occur, i.e., the resonance of *S*-factor at 260 keV simply cannot exist, although it is actually observed in experiment.

However, potentials No. 1 of Table 1 and No. 2 of Table 2, describing these two states (1st RS and 1st ES), are different – both have bound FS, but potential No. 2 has another BS, which is allowed bound 1st ES, and potential No. 1 leads to the resonance at 260 keV. We have previously considered cases when it was assumed that the potential parameters are explicitly dependent on Young diagrams {*f*}.[1–3] There was already a situation when there were different sets of such diagrams in BS and scattering processes, and therefore potentials could have different sets of parameters for these states. In particular, a similar situation occurred earlier only in one case, out of the 30 processes already considered,[1,2,17,22] with *M*1 capture in the $n^{16}O$ system to the $^2S_{1/2}$ first ES of $^{17}O$ from the nonresonance $^2S_{1/2}$ scattering wave.[2] It was necessary to admit possibility of such a transition in order to correctly explain the total cross sections of the neutron capture on $^{16}O$ at thermal energies. It was not possible to describe any other thermal cross sections within the framework of MPCM used by us.

In this case, it could be noticed again that we do not have exact tables of products of Young diagrams for number of particles equal to 15 in the 1 + 14 channel, which we used earlier for systems of particles with $A \leq 8$.[23] Thus, more accurate analysis of the states according to Young diagrams can lead to their difference for scattering and BS.

A similar explanation in this case allows us to assume presence of the $M1$ transition between "identical" $^2S_{1/2}$ or $^4D_{1/2}$ states. Therefore, here we will use different potentials of the same $JLS$ for BS and scattering processes.

However, transition No. 3 from Table 10, discussed above, does not allow us to correctly describe first peak of the $S$-factor, since it leads to a value about an order of magnitude smaller than the experiment.[50] The results of such a calculation with potentials No. 1 from Table 1 and No. 2 from Table 2 are shown in Fig. 5a with the red solid curve. The green dotted curve shows the results of calculating $S$-factor for processes No. 1 and 2 from Table 10 with zero $P$ wave potentials, and the blue dashed curve shows results for Process No. 3 from Table 10. Despite fact that at energies below the resonance, calculated curve proceeds almost according to the data of Ref. 50, the $S$-factor is 0.09(1) keV·b in the range of 30–100 keV, which is much higher (4–5 times) different results for the $S$-factor, a summary of which is given in Ref. 27.

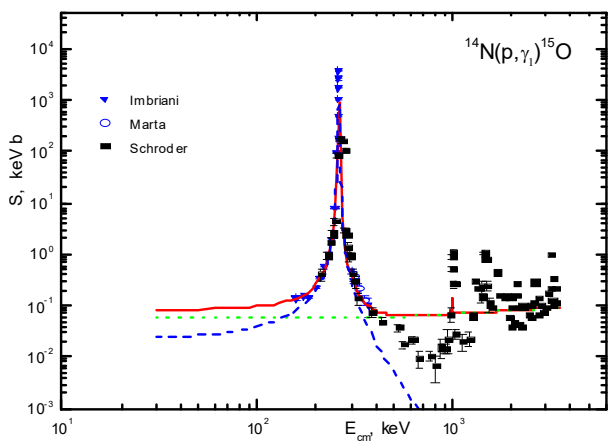

Fig. 5a. Astrophysical $S$-factor of the proton radiative capture on $^{14}$N to the first ES at 5.18 MeV of $^{15}$O. Experimental data are from Refs. 50,51,53. Curves are results of our calculations, described in the text.

In addition, second peak in the $S$-factor can be obtained only by considering $E2$ transition No. 4 from Table 10 for the $^2D_{3/2}$ waves with scattering potential No. 4 from Table 1. Moreover, for transition No. 5 from the $^2D_{5/2}$ waves (Table 10), the same potential No. 4 was used, obviously overstating calculation results. However, even in this case, as it is shown in Fig. 5a, calculated curve does not allow for the correct reproduce the magnitude of this resonance, since it leads to a peak an order of magnitude lower than the experimental data from Ref. 50. The experimentally observed minimum of $S$-factor between two resonances is also not reproduced. So, assumption that first ES is a $^2S_{1/2}$ wave does not look very conclusively.

*Second option for calculations*

First peak of the $S$-factor can be correctly described if the $^4D_{1/2}$ states are used for both the first resonance state and the first ES, i.e., take potential No. 2 from Table 1 and No. 3 from Table 2 and consider transition No. 8 from Table 10. The results of such calculation, taking into account $E1$ transitions No. 6,7 from the $P$ scattering waves (green dotted curve) to this level, are shown in Fig. 5b by the red solid curve. The blue dashed curve shows results for resonance $M1$ capture from all $D$ waves, i.e., processes No. 8,9 of Table 10. Only in such a case, resonance at 987 keV can be described at $M1$ transition (transition No. 9 from Table 10). It is enough to assume that this resonance is in the $^{2+4}D_{3/2}$ wave with potential No. 4 from Table 1.

As a result, it could be seen from calculated $S$-factor shown in Fig. 5b, its magnitude has a resonance at energy of 260 keV (transition No. 8) and 987 keV (transition No. 9) from Table 10, and accurately describe experimental data that are taken from Refs. 50,51,53. However, it should be noted that we consider second resonance at 987 keV, energy of which is taken from the review.[26] At the same time, the experimental data[51] have a maximum at 997 keV and do not agree with data[26] on resonance energy. These data have measurements of cross sections at 987 keV, but they lead to S-factor being approximately two times less than at 997 keV.

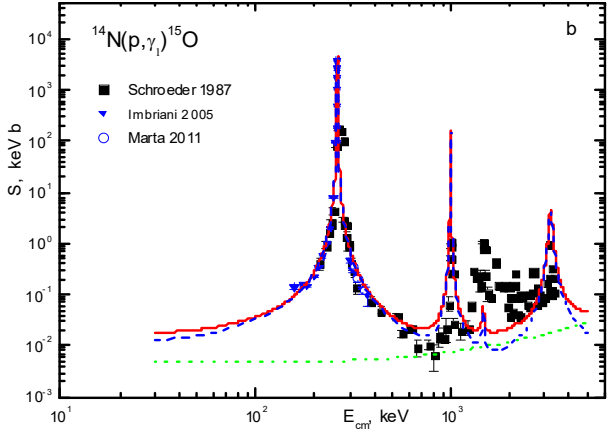

Fig. 5b. Astrophysical *S*-factor of the proton radiative capture on $^{14}$N to the first ES at 5.18 MeV of $^{15}$O. Notations the same as in Fig. 5a.

In these calculations, resonances from other *D* scattering waves are observed, the potentials of which are listed in Table 1, and the results are shown in Fig. 5b. As can be seen from Fig. 5b, contribution of the second resonance from the $^4D_{1/2}$ wave at 1447 keV is very small, and from second resonance the $^4D_{3/2}$ wave at 2187 keV is practically not observed. However, the last, fifth resonance at 3211 keV is clearly visible against background of existing maximum of experimental data at this energy.

The calculated *S*-factor shown in Fig. 5b at energy range of 30–60 keV has a value of 0.020(2) keV·b, which can be considered as its value at zero energy, and at 30 keV its value is equal to 0.018 keV·b. In Ref. 27, 0.033(25) keV·b was obtained for this *S*-factor, and 0.010(3) keV·b was proposed in Ref. 50. As can be seen from Fig. 5b, our results comparatively well describe the experimental data and the *S*-factor at zero energy in general is consistent with results of other works, which, on average, give 0.021 keV·b.

Thus, in this reaction, on basis of the approach used, it is quite possible to correctly describe the *S*-factors for transitions to the first ES in the $^4D_{1/2}$ wave of $^{15}$O for the $p^{14}$N channel from the first two and fifth resonance *D* scattering waves. It is difficult to call the reason why the third and fourth resonances are not described. In particular, the fourth resonance simply absent, and the third one is too small. However, it should be noted that experimental data from Ref. 51 at energies of approximately above 1.2 MeV in c.m. with the capture to the 1$^{st}$ES–3$^{rd}$ES indicate only the upper limit of values of the astrophysical *S*-factor.

Recall that potentials of *D* waves were used for the BS and scattering states, and this differs from results, for example, of Ref. 27 and the comparisons given in it with other works. It is usually considered that *S* waves correspond to these two states.[27,47,48,50,51,53] However, as it was evident from all of the results above, a more correct description of the *S*-factors in the resonance range is obtained under assumption that these resonances correspond to the *D* scattering waves.

Furthermore, if the real potential of the GS No. 1 and the 3$^{rd}$ES No. 5 from Table 2 is used for $P_{1/2,3/2}$ waves, then overall picture of description of the *S*-factor does not change, only its value at 30 keV becomes 0.015 keV In energy range of 30–60 keV, it varies in the interval of 0.017(2) keV·b.

### 4.4. *Astrophysical S-factor of the capture to the third ES*

Now let us consider radiative proton capture on $^{14}$N to the third excited state of $^{15}$O. Previously we considered this option in Ref. 6. This state has total moment $J = 3/2^-$ and can relate either to $^{2+4}P_{3/2}$ waves or to the $^{4}F_{3/2}$ wave. As usual, at first, we consider main possible transitions to such state and ways of averaging cross sections under such $E1$ or $M1$ transitions.

### 4.4.1. *Transitions at the capture to the 3$^{rd}$ES*

$E1$ transitions to the third mixed by spin $^{2+4}P_{3/2}$ excited state are possible

$$^{2}S_{1/2} \to\, ^{2}P_{3/2}$$
$$^{4}S_{3/2} \to\, ^{4}P_{3/2}.$$

For such transitions, as before, we have

$$\sigma(E1) = \sigma(^{2}S_{1/2} \to\, ^{2}P_{3/2}) + \sigma(^{4}S_{3/2} \to\, ^{4}P_{3/2}).$$

In addition, $E1$ processes are possible.

$$^{4}D_{1/2} \to\, ^{4}P_{3/2};\ ^{4}D_{3/2} \to\, ^{4}P_{3/2};\ ^{4}D_{5/2} \to\, ^{4}P_{3/2}$$

and

$$^{2}D_{3/2} \to\, ^{2}P_{3/2};\ ^{2}D_{5/2} \to\, ^{2}P_{3/2}.$$

When averaging states with different spin for such $E1$ transitions, we obtain

$$\sigma(E1) = \sigma(^{4}D_{1/2} \to\, ^{4}P_{3/2}) + [\sigma(^{2}D_{3/2} \to\, ^{2}P_{3/2}) + \sigma(^{4}D_{3/2} \to\, ^{4}P_{3/2})]/2 +$$
$$+ [\sigma(^{2}D_{5/2} \to\, ^{2}P_{3/2}) + \sigma(^{4}D_{5/2} \to\, ^{4}P_{3/2})]/2.$$

Furthermore, the following $M1$ transitions are possible

$$^{2}P_{1/2} \to\, ^{2}P_{3/2};\ ^{2}P_{3/2} \to\, ^{2}P_{3/2}$$
$$^{4}P_{1/2} \to\, ^{4}P_{3/2};\ ^{4}P_{3/2} \to\, ^{4}P_{3/2};\ ^{4}P_{5/2} \to\, ^{4}P_{3/2}.$$

They also require averaging over transitions between mixed states.

$$\sigma(M1) = \sigma(^{4}P_{5/2} \to\, ^{4}P_{3/2}) + [\sigma(^{2}P_{1/2} \to\, ^{2}P_{3/2}) + \sigma(^{4}P_{1/2} \to\, ^{4}P_{3/2})]/2 +$$
$$+ [\sigma(^{2}P_{3/2} \to\, ^{2}P_{3/2}) + \sigma(^{4}P_{3/2} \to\, ^{4}P_{3/2})]/2.$$

Here, as usual, it is considered that cross sections

$$\sigma(M1) = [\sigma(^{2}P_{3/2} \to\, ^{2}P_{3/2}) + \sigma(^{4}P_{3/2} \to\, ^{4}P_{3/2})]/2$$

are equal to zero at similar potentials of continuous and discrete spectra. As scattering potentials, you can use parameters of potential of the 3$^{rd}$ES No. 5 from Table 2, which lead to almost zero scattering phase shifts.

Now we give summary for Table 11 of all $E1$ and $M1$ transitions to the 3$^{rd}$ES considered above and coefficients $P^2$ in expressions for total cross sections. Here it is taken into account that the 3$^{rd}$ES can be not only the $^{2+4}P_{3/2}$ wave, but also the $^4F_{3/2}$ state; therefore, transitions for such a variant of orbital moment are given in Table 11 under No. 11–No. 15.

Table 11. Coefficients $P^2$ for total cross sections at transitions to the 3$^{rd}$ES

| No. | $[^{(2S+1)}L_J]_i$ | Type of transition | $[^{(2S+1)}L_J]_f$ | $P^2$ |
|---|---|---|---|---|
| 1 | $^2S_{1/2}$ | $E1$ | $^2P_{3/2}$ | 4 |
| 2 | $^4S_{3/2}$ | $E1$ | $^4P_{3/2}$ | 4 |
| 3 | $^4D_{1/2}$ | $E1$ | $^4P_{3/2}$ | 2/5 |
| 4 | $^4D_{3/2}$ | $E1$ | $^4P_{3/2}$ | 8/5 |
| 5 | $^2D_{3/2}$ | $E1$ | $^2P_{3/2}$ | 4/5 |
| 6 | $^4D_{5/2}$ | $E1$ | $^4P_{3/2}$ | 126/25 |
| 7 | $^2D_{5/2}$ | $E1$ | $^2P_{3/2}$ | 36/5 |
| 8 | $^2P_{1/2}$ | $M1$ | $^2P_{3/2}$ | 4/3 |
| 9 | $^4P_{1/2}$ | $M1$ | $^4P_{3/2}$ | 10/3 |
| 10 | $^4P_{5/2}$ | $M1$ | $^4P_{3/2}$ | 18/5 |
| 11 | $^4D_{1/2}$ | $E1$ | $^4F_{3/2}$ | 18/5 |
| 12 | $^4D_{3/2}$ | $E1$ | $^4F_{3/2}$ | 36/25 |
| 13 | $^4D_{5/2}$ | $E1$ | $^4F_{3/2}$ | 18/125 |
| 14 | $^4F_{5/2}$ | $M1$ | $^4F_{3/2}$ | 48/5 |

This Table was published in our work Ref. 6.

In this case, for the second option of the 3$^{rd}$ES potential, the transition from $^4F_{3/2}$ scattering wave is not taken into account, since in continuous spectrum the same potential is used as for BS and matrix element of such a transition is zero.

### 4.4.2. *Astrophysical S-factor of the capture to the 3$^{rd}$ ES*

Let us consider two options for calculating astrophysical $S$-factor $p^{14}$N capture for the third ES of $^{15}$O nucleus.

*First option of calculations*

With potentials given in Tables 1, 2, resonances in $D$ waves and 3$^{rd}$ES of $^{2+4}P_{3/2}$, S-factor calculations for $p^{14}$N radiative capture were performed at energies up to 1 MeV, taking into account the first 10 transitions listed in Table 11. Furthermore, the green dotted curve in Fig. 6a shows the results of the $E1$ capture from S scattering waves – transitions No. 1,2 in Table 11. The blue dashed curve shows the results for the $E1$ capture from all $D$ scattering waves with potentials No. 2 and No. 4 from Table 1 – processes No. 3–No. 7 in Table 11. The green dash-dotted line is the $M1$ processes from $P$ scattering waves – No. 8,9,10 in Table 11 with potentials of zero depth No. 9 in Table 1. The red continuous curve is the total result for the astrophysical $S$-factor of the capture process to the 3$^{rd}$ES.

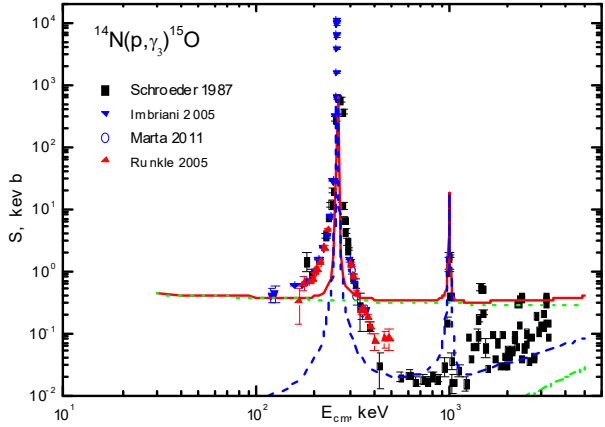

Fig. 6a. Astrophysical *S*-factor of the proton radiative capture on $^{14}$N to the 3$^{rd}$ES at 6.17 MeV of $^{15}$O. Experimental data are from Refs. 50–53. Curves are the results of our calculations, described in the text.

For nonresonance *D* waves, potential No. 8 from Table 1 was used, for nonresonance *S* potentials No. 7 from Table 1, the depth of *P* potentials was assumed to be zero, and the potential of the 3$^{rd}$ES is No. 5 in Table 2. Parameters of the 3$^{rd}$ES No. 5 from Table 2 were refined to correctly describe magnitude of experimental *S*-factor at the lowest energies of 119–124 keV. At these points, *S*-factor has values of 0.4(1) and 0.43(12) keV·b. And as can be seen from Fig. 6a, these experimental points are well described by calculated S-factor. Note that resonances above 1 MeV were not taken into account here.

As a result, as it can be seen from the calculated *S*-factor shown in Fig. 6a, its value has a resonance at energy of 260 keV (transition No. 3 from Table 11) and 987 keV (transitions No. 4,5), and positions of experimental resonances taken from Refs. 50–53. The first maximum agrees well with data of Ref. 51, and value of the second is somewhat larger than results of measurements from the same work Ref. 51. Note that in Ref. 51 the second maximum of the *S*-factor is observed at energy of 992 keV, i.e., it turns out to be shifted toward high energies by 5 keV relative to data from Refs. 26,28.

However, calculated minimum between the resonances, which is totally determined by the *S*-factor at the capture to the 3$^{rd}$ES from *S* scattering waves, turned out to be noticeably larger than experimental results. As it can be seen from Fig. 6a, it is not possible to describe *S*-factor on the basis of the first resonance, which is totally determined by its width. At the same time, theoretical calculations of width Γ with resonance potential No. 2 from Table 1 accurately reproduce experimental value.[26,28]

Note that only with capture to this odd ES, it is not possible to correctly describe the width of the *S*-factor for the first resonance. In all the previous cases discussed above with the same resonance $^4D_{1/2}$ potential, this problem was not observed. The calculated *S*-factor shown in Fig. 6a at energy range of 30–200 keV has a value of 0.40(3) keV·b, which can apparently be considered as its value at zero energy. The error is determined by the averaging of *S*-factor in specified energy range. The *S*-factor value at low energies is totally determined by the capture from *S* waves.

We present in Table 12 for comparison some experimental values for the *S*-factor at zero energy from other works.

Table 12. *S*-factor values at zero energy obtained in different works

| Refs. | 27 | 50 | 52 | 31 | 46 | Average value |
|---|---|---|---|---|---|---|
| $S(0)$, keV·b | 0.117(37) | 0.08(3) | 0.04(1) | 0.06(2) | 0.13(2) | 0.085 |

As it can be seen from this table, value of the calculated total *S*-factor obtained above in stabilization range of its values is noticeably greater than results of these works, which lead to interval 0.03–0.15 keV·b with average over interval 0.09(6) keV·b.

Furthermore, we note that *S*-factor value turned out to be less sensitive to shape of the $^{2,4}S$ wave potential. The use of *S* potentials, for example, with parameters 420.0 MeV and $\alpha = 1.0$ fm$^{-2}$, which lead to scattering phase shift of 180(1) degrees in energy range up to 1 MeV, changes value of the *S*-factor at 30–200 keV only to 0.44(2) keV·b. Changes in parameters of potential for *P* waves within relatively large limits, but so that it leads to scattering phase shifts close to zero, also have almost no effect on shape and magnitude of the *S*-factor. Once again, we note that in order to determine parameters of these potentials more unambiguously, results of phase shift analysis of the $p^{14}$N scattering, which are currently absent, are required. If, for potential of the $P_{1/2}$ scattering wave, we use potential of the GS the capture *S*-factor from *P* waves will decrease somewhat, without affecting its general shape.

So, in this reaction, based on approach used and *P* wave in the 3$^{rd}$ES, it is not possible to correctly describe behavior of the experimental *S*-factor for transitions to the 3$^{rd}$ES of $^{15}$O in the $p^{14}$N channel. It is possible to describe only its general form, although details of calculations, for example, interval between resonances, and its values at zero energy are not reproduced. As in the previous paper, potentials of the *D* waves were used for resonance scattering states, which differs from results, for example, in Ref. 27 and comparisons given in it with other works.

*Second option of calculations*

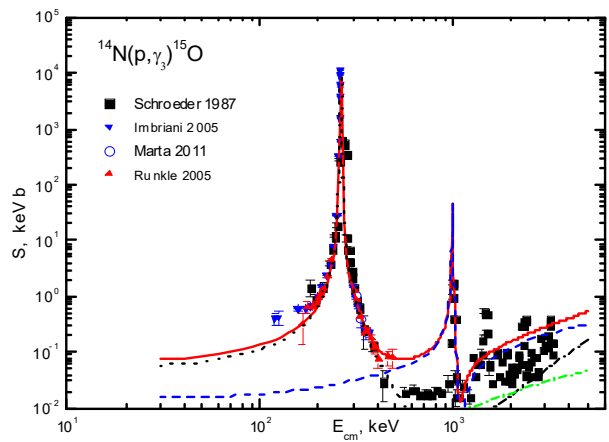

Fig. 6b. Astrophysical *S*-factor of the proton radiative capture on $^{14}$N to the 3$^{rd}$ES at 6.17 MeV of $^{15}$O. Notations as in Fig. 6a.

There is, however, another option of calculations given above. If we at first do not use $^{2+4}P_{3/2}$ wave for the 3$^{rd}$ES, but the $^{4}F_{3/2}$ wave, possible transitions are listed in second part of Table 11 under No. 11–14, and potential of the 3$^{rd}$ES is given in Table 2 under No. 6. All scattering potentials are listed in Table 1 and do not differ from previous case. Both resonances are still in the *D* waves, and for *F* scattering waves the *P* potentials No. 9 from Table 1 are used. For the potential of the nonresonance $D_{5/2}$ wave, interaction No. 8 from Table 1 is used, and results of calculations are shown in Fig. 6b.

The green dash-dotted line is *E*1 transition from $D_{5/2}$ wave (No. 13 from Table 11), black dash-dotted curve is *M*1 transition from *F* wave (No. 14), black dotted line is *E*1 – the transition from $D_{1/2}$ wave (No. 11), blue dashed curve is *E*1 transition from $D_{3/2}$ wave (No. 12), red solid curve is sum of all transitions. As it can be seen from these results, the *S*-factor at low energies in the range of 30–60 keV has a value of 0.08(1) keV·b and agrees well with average value given in Table 12.

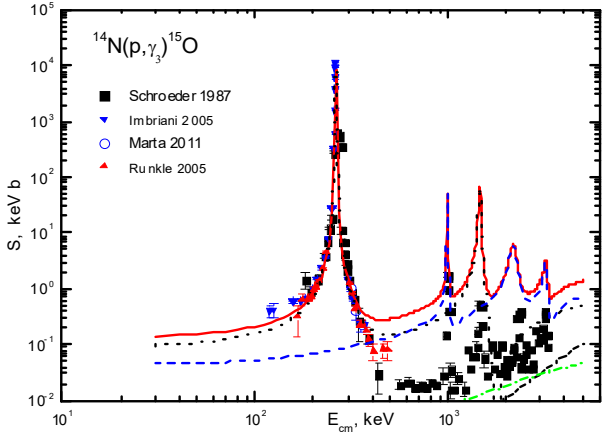

Fig. 6c. Astrophysical *S*-factor of the proton radiative capture on $^{14}$N to the 3$^{rd}$ES at 6.17 MeV of $^{15}$O. Notations as in Fig. 6a.

In contrast to results presented in Fig. 6a, width of the first resonance measured in Ref. 52 is now correctly described. At maximum, calculation results are in good agreement with new measurements Ref. 50, which are an order of magnitude larger than results of Ref. 51. It is possible to correctly describe data of Ref. 52 in the interresonance energy range, which, at energies above 400 keV, are located noticeably higher than earlier results.[51] In energy region before the first resonance, calculation results agree with new experimental data from Ref. 52, which differ noticeably from measurements.[50]

The calculated value of the second resonance is order of a magnitude higher than data of Ref. 51, but in this area there are no measurements similar to result of Ref. 50, which can also have a large value, as in case of the first resonance here. With results of Ref. 52 our calculations are in good agreement with both in first resonance range and in interresonance energy range, i.e., wherever measurements of the *S*-factor are performed in Ref. 52. If we consider available experimental data from Ref. 51, then from Fig. 6b one can see correct description of a minimum after the second resonance.

From these results it can be seen that it is preferable to use *F* wave as the 3$^{rd}$ES. In fact, the 3$^{rd}$ES is a mixture of three states – $^2P_{3/2}$, $^4P_{3/2}$ and $^4F_{3/2}$ and the third of them seems to be the most likely to happen. In addition, potential No. 6 of Table 2 describes AC more correctly, values of which for all ES are given in Table 6.

Now we take into account three new resonances at energies from 1447 to 3211 keV with parameters from Table 1 and carry out calculations up to 5 MeV. The results of these calculations are shown in Fig. 6c – green dash-dotted curve is the *E*1 transition from $D_{5/2}$ wave (No. 13), black dash-dotted curve is the M1 transition from F wave (No. 14), black dotted line is the *E*1 transition from two $D_{1/2}$ waves (No. 11), blue dashed curve is the *E*1 transition from three $D_{3/2}$ waves (No. 12), red solid curve is the sum of all transitions. From these results, we see that *S*-factor in the range of 30–60 keV increased its value up to 0.15(1) keV·b and no longer agrees with average value given in Table 12.

Description of three new resonances turns out to be noticeably worse than it was when considering previous capture reactions to other ES. In contrast to previous processes, here for the first time we have a state with negative parity. All previous BS had positive parity and, apparently, related to the *D* waves. So, results presented in Fig. 6b and especially Fig. 6c at the radiative capture to the negative parity BS describe available experimental data noticeably worse than they did in previous cases.

### 4.5. *Astrophysical S-factor for capture to the GS*

Here we consider two options of calculations with different GS potentials of $^{15}$O in the $p^{14}$N channel. In the first case, potential of $P_{1/2}$ wave is used for the GS, and in the second case we will take into account possible excitation of the $^{14}$N cluster. Previously we considered this option in Ref. 7.

### 4.5.1. *Transitions at the capture to the GS*

If we consider the GS of $^{15}$O in the $p^{14}$N channel with $^{14}$N cluster in the ground state, then it turns out to be mixed by spin $^{2+4}P_{1/2}$. Therefore, it is possible to consider $E1$ transition from $^{2}S_{1/2}$ scattering wave to the doublet part of $^{2}P_{1/2}$ wave function of the GS of $^{15}$O.

$$^{2}S_{1/2} \rightarrow {}^{2}P_{1/2}.$$

In addition, it is possible to consider $E1$ transition from $^{4}S_{3/2}$ scattering wave to the quartet part of $^{4}P_{1/2}$ wave function of the GS

$$^{4}S_{3/2} \rightarrow {}^{4}P_{1/2}.$$

The $E1$ transitions of the next form are also possible

$$^{2}D_{3/2} \rightarrow {}^{2}P_{1/2}$$
$$^{4}D_{3/2} \rightarrow {}^{4}P_{1/2}.$$
$$^{4}D_{1/2} \rightarrow {}^{4}P_{1/2}$$

In this case, $E1$ transitions take the place between $^{2+4}D_{3/2}$ mixed by spin scattering states and also mixed by spin $^{2+4}P_{1/2}$ GS. Therefore, the total cross section of capture process to the GS of $^{15}$O for $E1$ electromagnetic transitions from $^{2+4}D_{3/2}$ waves are represented as following combination of partial cross sections

$$\sigma(E1) = \left[\sigma(^{2}D_{3/2} \rightarrow {}^{2}P_{1/2}) + \sigma(^{4}D_{3/2} \rightarrow {}^{4}P_{1/2})\right]/2.$$

Here, as before, averaging over transitions with the same total moment, but different spin of channel is performed. In this case, total cross section of the process of capturing to the GS for electromagnetic $E1$ transitions are presented in form of following combination of partial sections

$$\sigma_0(E1) = \sigma(^{2}S_{1/2} \rightarrow {}^{2}P_{1/2}) + \sigma(^{4}S_{3/2} \rightarrow {}^{4}P_{1/2}) + \sigma(^{4}D_{1/2} \rightarrow {}^{4}P_{1/2}) +$$
$$+ \left[\sigma(^{2}D_{3/2} \rightarrow {}^{2}P_{1/2}) + \sigma(^{4}D_{3/2} \rightarrow {}^{4}P_{1/2})\right]/2$$

In addition, one can consider $M1$ transitions between $P$ waves of continuous spectrum and the $^{2+4}P_{1/2}$ GS of nucleus.

$$^{2}P_{1/2} \rightarrow {}^{2}P_{1/2}$$
$$^{2}P_{3/2} \rightarrow {}^{2}P_{1/2}$$
$$^{4}P_{1/2} \rightarrow {}^{4}P_{1/2}$$

$$^4P_{3/2} \rightarrow {^4P_{1/2}}.$$

Here we also make averaging over transitions between mixed states.

$$\sigma(M1) = [\sigma(^2P_{1/2} \rightarrow {^2P_{1/2}}) + \sigma(^4P_{1/2} \rightarrow {^4P_{1/2}})]/2 +$$
$$+ [\sigma(^2P_{3/2} \rightarrow {^2P_{1/2}}) + \sigma(^4P_{3/2} \rightarrow {^4P_{1/2}})]/2$$

If we use further for $^{2+4}P_{1/2}$ scattering states the same potential as for the GS, the $M1$ transitions between them become impossible, since their wave functions turn out to be orthogonal, and matrix elements of first term lead to zero sections.

Furthermore, we present a summary Table 13 of all $E1$ and $M1$ transitions to the GS considered above without taking into account transitions from $^{2+3}P_{1/2}$ waves and coefficients $P^2$ in expressions for total cross sections given above.

Table 13. $P^2$ coefficients for transitions to the GS

| No. | $[^{(2S+1)}L_J]_i$ | Transition type | $[^{(2S+1)}L_J]_f$ | $P^2$ |
|---|---|---|---|---|
| 1 | $^2S_{1/2}$ | $E1$ | $^2P_{1/2}$ | 2 |
| 2 | $^4S_{3/2}$ | $E1$ | $^4P_{1/2}$ | 2 |
| 3 | $^4D_{1/2}$ | $E1$ | $^4P_{1/2}$ | 2 |
| 4 | $^2D_{3/2}$ | $E1$ | $^2P_{1/2}$ | 4 |
| 5 | $^4D_{3/2}$ | $E1$ | $^4P_{1/2}$ | 2 |
| 6 | $^2P_{3/2}$ | $M1$ | $^2P_{1/2}$ | 4/3 |
| 7 | $^4P_{3/2}$ | $M1$ | $^4P_{1/2}$ | 10/3 |

### 4.5.2. *Astrophysical S-factor of the capture to the GS*

Now we consider two options for calculated astrophysical $S$-factor of the proton capture on $^{14}$N to the ground state of $^{15}$O.

*First option of calculations*

Let us first give some experimental values for the $S$-factor at the capture to the GS of $^{15}$O, which are grouped in Table 14. In one of the most recent studies on behavior of astrophysical $S$-factor of the proton capture on $^{14}$N, including the capture to the GS,[54] the smallest of all 0.19(5) keV·b is given.

Table 14. $S$-factor values at zero energy, obtained in different works

| Refs. | 27 | 50 | 52 | 55 | 46 | 49 | 54 |
|---|---|---|---|---|---|---|---|
| $S(0)$, keV·b | 0.25(5) | 0.25(6) | 0.49(8) | 0.27(5) | 0.15(7) | 0.42(4) | 0.19(5) |

This Table was published in our work Ref. 7.

From these results, we obtain range of values of 0.08–0.57 keV·b, which gives an average over interval value of 0.33(25) keV·b. If we average 7 given values, we get an average value of 0.29 keV·b. As it can be seen, scatter of values and their errors are quite large, which currently does not allow us to make an unambiguous conclusion about value of *S*-factor at zero energy. The experimental data for the astrophysical *S*-factor at energies up to 1 MeV are taken from Refs. 50–52,56 and are shown in Fig. 7 by different points.

From Fig. 7, it is clear that only last two points from Ref. 50 undergo rise at the lowest energies, although it is difficult to say up to what values this rise can lead – experimental errors are too large. This is the reason for such a large scatter of the *S*-factor values at zero energy, obtained in works given above. The errors of these points at 119 and 124 keV give a lower *S*-factor value of 0.01 keV·b. This value is below than previous point for 157 keV, which is located at 0.015(4) keV·b.

If you do not take into account rise of these points, then *S*-factor value, at least, its lower value, can be approximately at 0.01 keV·b, and upper value, taking into account these two points, can reach a value of 0.5 and even 1.0 keV·b. In fact, ambiguity of such results reaches two orders of magnitude and only that, simply "requires" new experimental measurements to be carried out with higher accuracy.[57] From available experimental measurements, all possible information has already been extracted, which, as it was shown above, leads to very large *S*-factor ambiguities in the range of low and zero energy.

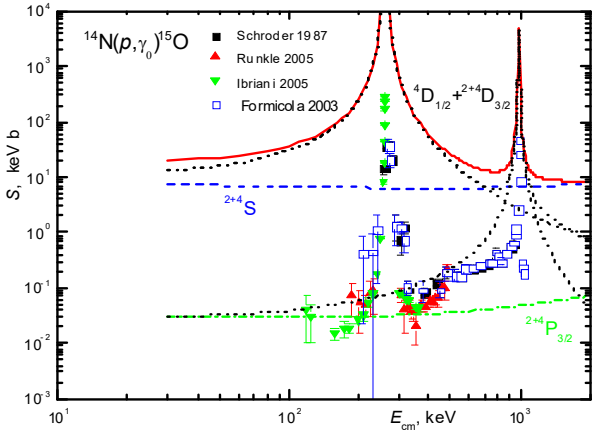

Fig. 7. Astrophysical *S*-factor of the proton radiative capture on $^{14}$N to the GS of $^{15}$O. Experimental data are from works Refs. 50–52,56. Curves are the results of our calculations, described in the text.

Furthermore, with potentials given in Table 1 and Table 2, which have resonances in the *D* waves of scattering, nonresonance *S* and *P* waves, calculations of the astrophysical *S*-factor for the proton radiative capture on $^{14}$N to the GS in the $^{2+4}P_{1/2}$ waves were performed at energies up to 1 MeV. All 7 transitions listed in Table 13 were taken into account in calculations. As it was already mentioned, transitions between $^{2+4}P_{1/2}$ states lead to zero matrix elements, since they used potential No. 1 from Table 2, both in the GS and in scattering processes.

In Fig. 7, black dotted curves show results of the *E*1 capture from the resonance *D* scattering waves with potentials No. 2 and No. 4 from Table 1. Blue dashed curve shows results for the *E*1 capture from two nonresonance *S* waves with same potential No. 7 from Table 1. The green dash-dotted curve shows *M*1 capture from $^{2+4}P_{3/2}$ waves with potential of zero depth No. 9 from Table 1. The red solid curve shows total result for astrophysical *S*-factor of capture process to the GS, taking into account capture from all scattering waves considered above.

As a result, as it can be seen from Fig. 7, the value of calculated *S*-factor has a resonance at energy of 260 keV (transition No. 3 from Table 13) and 987 keV (transitions No. 4,5), however, both maxima are much higher than the results of experimental work. At the same time, *S*-factor of capture from *P*-scattering waves is in error region of last two points of work Ref. 50. Magnitude of calculated *S*-factor in the first resonance reaches 3.7 10$^6$ keV·b, in the second 5.4 10$^3$ keV·b. Moreover, such increase in values has exactly S-factor in capture of scattering from D waves, which is shown in Fig. 7 by black dotted curves. At energy of 30 keV, astrophysical S-capture factor from S waves has calculated value of 7.5 keV·b, and for a P wave it is 0.03 keV·b – this value agrees well with experiment at low energies. Experimental values of the S factor in the resonances are respectively 293 keV·b at 259 keV Ref. 50 and 48 keV·b at 990 keV Ref. 56.

Potentials of S and P waves used in calculations lead to zero scattering phases, but if we slightly change their parameters, it does not lead to any significant changes in calculation results. For comparison, potential of S-waves of zero depth was used, which led to decrease in S-factor to magnitude of 5.9 keV·b. With arbitrary change in potential parameters of GS, even to a width of 12 fm$^{-2}$, which leads to dimensionless AC equal to 0.45(5), results of calculations still do not allow us to correctly describe the experimental S-factor. With minimum energy of 30 keV, the S-factor value will decrease only by two orders of magnitude with sharp rise in the same shape as shown by continuous curve in Fig. 7. The magnitude of the first maximum is also not described, although the second calculated maximum is lower than it is in experiment. As a result, it can be seen that any potentials used in the work do not allow us to correctly describe even order of the experimental S-factor.

Let us now consider possibility of describing the first resonance at 260 keV using resonance $^2S_{1/2}$ wave. For such scattering potential with FS, parameters No. 1 from Table 1 were used. For nonresonance potential, now $^4D_{1/2}$ scattering wave with FS one can use parameters No. 8. For the $^4S_{3/2}$ scattering wave, we use potential No. 7, and all other potentials do not change. In this case, calculations lead to the S-factor at energy of 30 keV, which is an order of magnitude greater than the results shown in Fig. 7 by the red solid curve. Also magnitude of the first resonance also slightly increases. From these results it can be seen that usage of this option of potential parameters also does not lead to improvement in description of experimental data in entire energy range.

Since we do not have ability to accurately classify states according to Young diagrams, it can be assumed that number of bound forbidden states in the GS is not correctly determined. Therefore, it is possible to construct two other options of GS potentials with one and two FS, so that their ACs are approximately the same. In this case, parameters 209.60479 MeV and 0.2 fm for AC = 6.1 were obtained for the GS potential with one FS. For the second option with two FS, parameters 381.3004 MeV and 0.2 fm were obtained with AC = 8.3. However, usage of such potentials led in both cases to an increase in S-factor at 30 keV to approximately 30 keV b.

As a result, we can conclude that, within framework of an approach we used, we could not find a definite explanation for experimental results obtained. For calculations of capture from the resonance D waves, the same potentials were used, which earlier allowed us to explain capture processes for all five ES. Therefore, reason for this discrepancy between results of calculation and experiment can only be in the GS potential. Some reason strongly reduces amplitude of WF of the GS,[9] which leads to experimentally observed values of the astrophysical S-factor.

*Second option of calculations*

However, there is a way to explain order of magnitude of the experimental S-factor at low energies. Moreover, two-body MPCM, described particularly in Refs. 1,2,16–18,22, is still used, with one additional assumption. Namely, it is sufficient to assume that $^{14}$N nucleus can be inside $^{15}$O in the fifth excited state of $^{14}$N* with the excitation energy of 5.69 MeV and a moment of $J^\pi = 1$. In this case, it is possible to almost correctly reproduce available experimental data of the proton capture on $^{14}$N to the GS. Under this assumption, wave function of relative motion of clusters can be a quartet $^4D_{1/2}$ wave, and binding energy of the $p^{14}$N* system now equals $E_b(p^{14}$N*$) = E_b(p^{14}$N$) + E_x(^{14}$N*$) = -7.2971 + 5.6914 = -1.6057$ MeV Ref. 26. Thus, the GS of $^{15}$O is now considered in two-particle $p^{14}$N* channel with excited $^{14}$N* cluster.

In course of this reaction, immediately after the proton capture on $^{14}$N to the GS of $^{15}$O, γ-quanta with energy of 7.2971 MeV is emitted, which is recorded in experiment, and at the same time $^{14}$N cluster passes into the excited state. In other words, after the emission of γ-quanta the GS of $^{15}$O passes into stationary $^4D_{1/2}$ state of the $p^{14}$N* channel with excited $^{14}$N* cluster. Part of binding energy of the $p^{14}$N system goes to excitation of the $^{14}$N cluster, so that relative energy of new $p^{14}$N* system becomes -1.6057 MeV. As a result, we assume that wave function of this stationary $^4D_{1/2}$ state can be used to calculate the astrophysical $S$-factor.

In other words, it is possible to represent the WF of the GS of $^{15}$O after emitting γ-quanta as superposition of functions over all possible excitations of the $^{14}$N cluster,

$$\Psi(GS) = \alpha\psi_0[p^{14}N] + \beta\psi_1[p^{14}N^1] + \gamma\psi_2[p^{14}N^2] + \ldots,$$

and the coefficients α,β,γ etc. are probability of such states. As a result, WF $\psi_0[p^{14}N]$ determines the GS of $^{15}$O with the $^{14}$N cluster in the ground state; function $\psi_1[p^{14}N^1]$ determines the GS with the $^{14}$N$^1$ cluster in the first possible excited state, etc.

In all 30 processes of the radiative capture we considered earlier (see, for example, Refs. 1,2,16–19,22), main contribution was made only by the first member of this expansion, i.e., the coefficient α was predominant. In this case, we first encountered, apparently, presence of the excited $^{14}$N* nucleus in the $p^{14}$N cluster system, and probability of just such channel turned out to be predominant. Note that the GS in this channel that may be the $^2S_{1/2}$ wave, however, set of possible transitions to such a state was not able to correctly describe available experimental data.

Furthermore, parameters of the Gaussian GS potential of $^{15}$O in the $p^{14}$N* channel were obtained, which are listed in Table 15. In these calculations, we still believe that all scattering resonances are in $D$ waves with the FS, for nonresonance $^4S_{3/2}$ wave with FS, parameters are used that lead to the scattering phase shifts close to zero, and potentials in $P$ waves have zero depth – see Table 1. The resulting GS potential leads to charge radius of $^{15}$O nucleus equal to 2.9 fm, and for the radius of $^{14}$N* previous value of 2.560(11) fm was used,[41] as in the case of $^{14}$N nucleus. Since we could not find information on asymptotic constant in such GS, potential parameters were selected to best describe $S$-factor at zero energy, which is entirely determined by the $E1$ transitions from $P$ scattering waves with zero potential from Table 1.

Table 15 Parameters of the $p^{14}\text{N}^*$ GS potential. The fifth and sixth columns show the binding energy and its AC.

| Bound $p^{14}\text{N}^*$ level, which is matched to the GS | $V_0$, MeV | $\alpha$, fm$^{-2}$ | Binding energy, MeV | AC $C_w$ |
|---|---|---|---|---|
| $^4D_{1/2}$ | 134.246725 | 0.09 | -1.60570 | 3.0(1) |

This Table was published in our work Ref. 7.

When considering process of the proton capture on $^{14}\text{N}$ to such GS, the radiative transitions between scattering states shown below in Table 16 were taken into account.

Table 16. Radiative transitions at the proton capture on $^{14}\text{N}$ to the GS of $^{15}\text{O}$ in the $p^{14}\text{N}^*$ channel

| No. | $[^{(2S+1)}L_J]_i$ | Type of transition | $[^{(2S+1)}L_J]_f$ | $P^2$ |
|---|---|---|---|---|
| 1 | $^4P_{1/2}$ | E1 | $^4D_{1/2}$ | 2 |
| 2 | $^4P_{3/2}$ | E1 | $^4D_{1/2}$ | 2/5 |
| 3 | $^4S_{3/2}$ | E2 | $^4D_{1/2}$ | 2 |
| 4 | $^4D_{3/2}$ | E2 | $^4D_{1/2}$ | 2 |
| 5 | $^4D_{1/2}$ | M1 | $^4D_{1/2}$ | 3/2 |
| 6 | $^4D_{3/2}$ | M1 | $^4D_{1/2}$ | 6 |
| 7 | $^4F_{3/2}$ | E1 | $^4D_{1/2}$ | 12/5 |

This Table was published in our work Ref. 7, but without 7$^{th}$ transition.

Capture of the *M*1 type from the $^4D_{1/2}$ scattering wave to the $^4D_{1/2}$ bound state turns out to be possible, since different potentials are used for these states. Each of these potentials describes its own set of characteristics, which belongs to these states and there is no possibility to get unique potential for them. Reason for this may be, for example, in difference of Young diagrams for such states, as it was considered earlier in Refs. 3,17. In this case, situation may be similar, since above consideration of states according to Young diagrams is only qualitative, due to lack of exact product tables of these diagrams as it was in Ref. 23.

Results of calculations with such potentials and above considered transitions are shown in Fig. 8. Blue dashed curve are capture processes No. 4,5,6 from Table 16, the green dash-dotted curve is capture No. 1,2 from Table 16 and the red solid curve is their sum. Capture No. 3 is shown at the bottom by black dotted curve – it makes almost no contribution to the total *S*-factor. $^4D_{1/2} \rightarrow {}^4D_{1/2}$ transition is considered to be possible because different potentials from Table 1 and Table 15 were used for these states in continuous and discrete spectra. The last process No. 7 from Table 16 with potential No. 10 from Table 1 and the GS potential from Table 15 nevertheless leads to the peak of the curve in Fig. 8 at the resonance energy, but due to the small width its influence to the considered further reaction rate will be minimal.

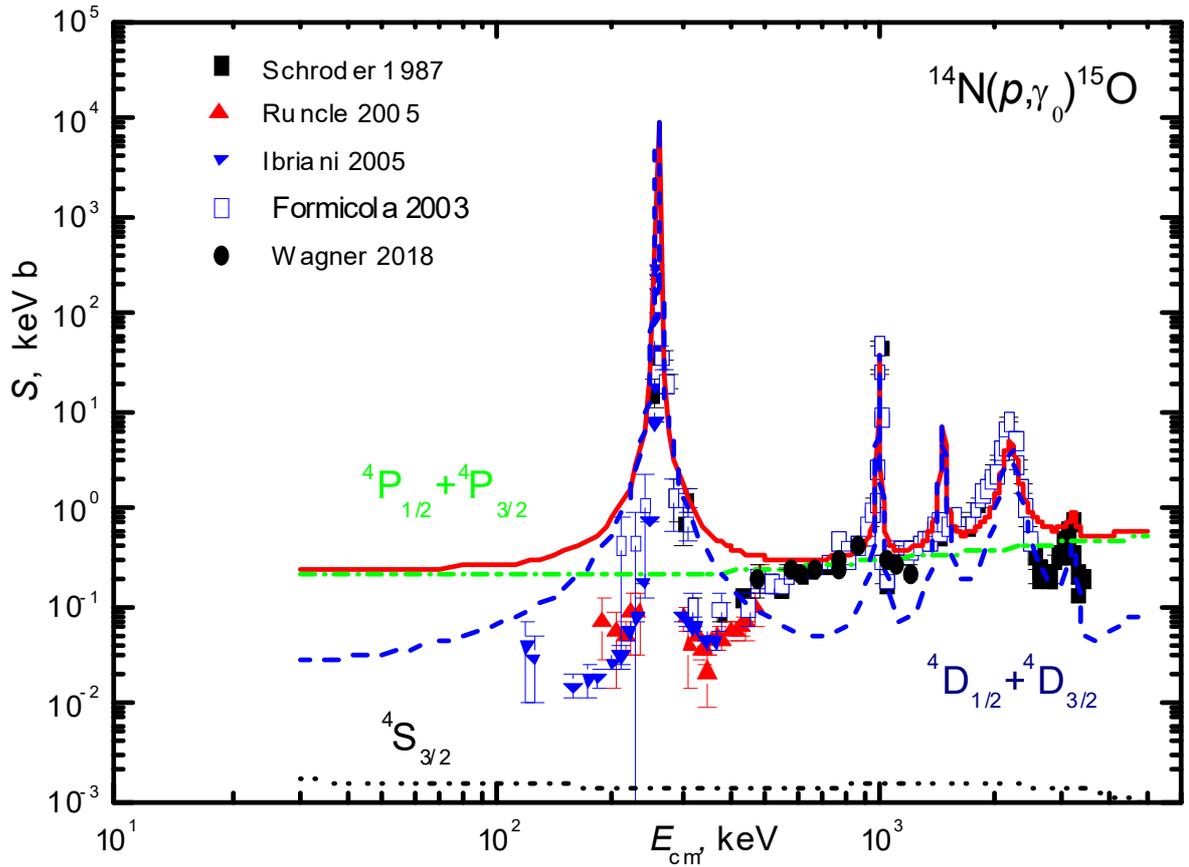

Fig.8. Astrophysical $S$-factor of the radiative capture to the GS of $^{15}$O, which is considered in the $p^{14}$N* channel. Experimental data are taken from Refs. 50–52,56. Curves – results of our calculations, which are described in the text.

At an energy of 30 keV, the value of $S$-factor turned out to be 0.24 keV·b, and in the energy range of 30–110 keV, this value is of 0.26(2) keV·b. As it can be seen from Fig. 8, calculated value of the first resonance turns out to be somewhat larger than experimental data, reaching a value of 9.5 10$^3$ keV·b, with an experimental value of 292 keV·b.[50] The width of the first resonance as a whole is correctly described, but in the range of lowest energies the calculated $S$-factor goes noticeably higher than data,[50,52] although its value in stabilization region of its values, which we consider to be zero energy, totally coincides with results of works listed in Table 14. This value is provided by the capture from $P$ scattering waves with potentials of zero depth, as shown in Fig. 8 by the green dash-dotted curve.

From Fig. 8 one can see a good description of value of the second maximum at 990 keV equal to 40 keV·b with an experimental value of 48 keV·b.[50] In interresonance energy range, the calculated $S$-factor is generally consistent with measurements,[54,56] especially in the second resonance region. Although rate of decrease of the calculated $S$-factor after the first resonance is not as strong as experimental measurements of well-known works Refs. 50–52,54,56 show.

Magnitude of the third resonance turns out to be an order of magnitude larger than available data, and the fourth resonance agrees well with measurement results.[56] As it can be seen from Fig. 8, experiment simply lacks data for the third resonance in the $^4D_{1/2}$ wave at 1447 keV in c.m. Although in the range of this resonance $S$-factor undergoes a slight rise, it is certain that at resonance energy cross section measurements were not performed.[51,56] The fifth resonance is also present, and its height is consistent with the data of Ref. 51, but minimum between it and previous resonance is described relatively poorly. This result is also due to the *S*-factor behavior in the case of *E*1 capture from *P* scattering waves – green dash-dotted curve in Fig. 8. Potential of these waves was assumed to be zero, but at 2–3 MeV this assumption is no longer correct.

In this version of calculations, question of parameter's uniqueness of *S* and *P* scattering wave potentials remains open. In absence of results of the phase shift analysis of the $p^{14}$N elastic scattering, even knowing number of BS in each of these partial waves, it is impossible to totally unambiguously construct such potentials. However, direct verification showed that results of calculations of the astrophysical *S*-factor weakly depend on parameters of these potentials, only the condition that partial scattering phase shift value close to zero is important. But still, to construct more correct potentials, we need results of the phase shift analysis of the $p^{14}$N elastic scattering.

Thus, assumption made by us about possibility of the excited state of the $^{14}$N$^*$ cluster at energy of 5.69 MeV with moment $J = 1^-$ in $^{15}$O makes it possible to describe experimental results of the astrophysical *S*-factor of the proton radiative capture on $^{14}$N at low energy range. Perhaps, here it is necessary to take into account not only this cluster channel in the considered excited $^{14}$N$^*$ state, but also other permissible excitations of $^{14}$N. However, it is already clear from these results that this particular channel with considered excitation of $^{14}$N nucleus is predominant and makes the greatest contribution to the astrophysical *S*-factor of the proton capture process on $^{14}$N to the GS of $^{15}$O.

## 5. Total *S*-factor and reaction rate

Now we can show all *S*-factors on the same figure, and summing them up, get the total *S*-factor for capturing to all considered BSs of $^{15}$O. These results are shown in Fig.9a,b, and sum for all transitions is shown by the red solid curve. It also presents, apparently, the most recent experimental data from Ref. 58, which lead to the value of *S*-factor at energy of 70–108 keV in interval 1.68–1.77 keV·b, with average value of 1.73(4) keV·b. For zero energy, Ref. 58 gives 1.74 ± 0.14 (stat) ± 0.14 (syst) keV·b. Our calculations in the energy range of 30–100 keV lead to the *S*-factor of 1.78–1.85 keV·b with an average of 1.82(4) keV·b. In addition, the estimated error may appear to be about 5%, i.e., about 0.1 keV·b. Only two first resonances were taken into account for *S*-factors and are shown in Fig. 9a, because just those most strongly influence to the reaction rate.

For a better presentation of measurements and calculation results at the lowest energies Fig. 9b, which covers energy range up to 250 keV is shown. As it can be seen from Fig. 9, there is a good agreement between new data of the total *S*-factor from Ref. 58 (blue dots) and results of our calculations at the lowest energies. Fig. 9 also shows the results of Ref. 46, which are shown by the green triangles. Our calculations also agree well with them in low-energy range. Although, as it can be seen in Fig. 9b, at energies above 150–200 keV, a more rapid growth of calculated *S*-factor is observed compared to results of Refs. 46,58, which may indicate to a greater width of resonance obtained in the calculations.

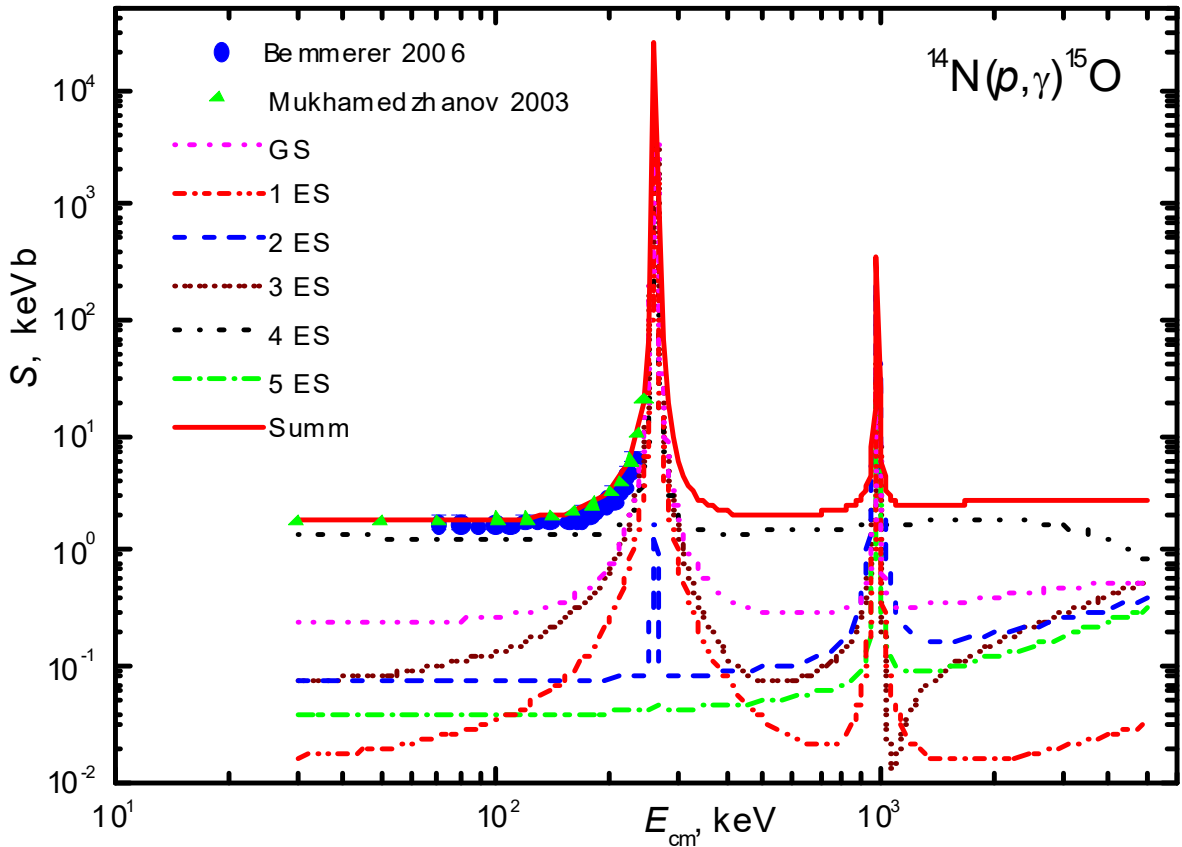

Fig. 9a. Astrophysical S-factor of the radiative capture to all BS of $^{15}$O. Curves – results of our calculations, described in the text.

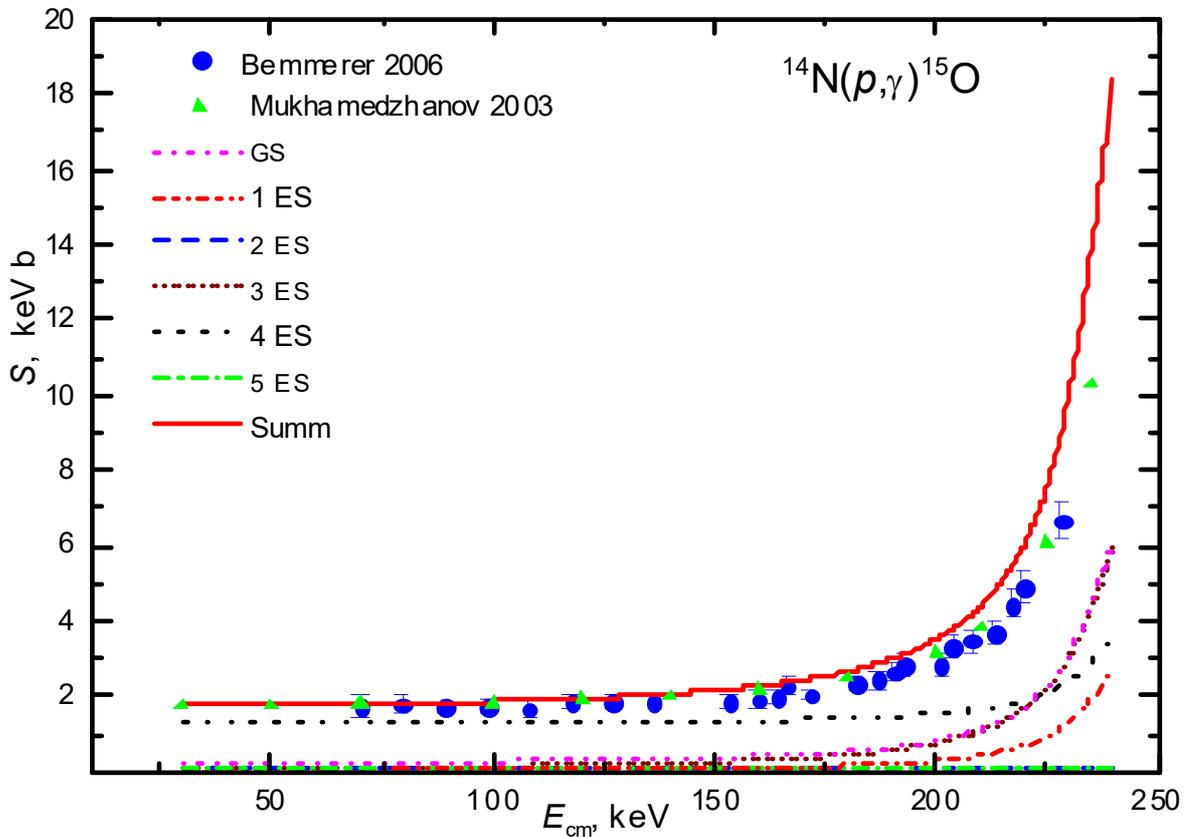

Fig. 9b. Astrophysical S-factor of the radiative capture to all BS of $^{15}$O. Curves – results of our calculations, described in the text.

Based on *S*-factors obtained for capturing to all BSs shown in Fig. 9, it is possible to calculate reaction rates that correspond to such captures — they are shown in Fig. 10 in temperature range from 0.01 to 10 $T_9$. To calculate the reaction rate, the *S*-factors were used in the range from 10 keV to 5 MeV with a step of 10 keV. Since, the calculation of *S*-factor for all transitions started from 30 keV, then for energies of 10 and 20 keV, the average calculated *S*-factor was taken for capture to each BS. Total capture rate to all BSs is shown in Fig. 10 by the blue solid curve and was calculated for 1000 points with a step of 0.01 $T_9$ in indicated temperature range.

Fig. 10 shows a comparison with results of other works. In particular, results obtained for reaction rate in Ref. 27, which are represented by the red dots, are shown. They do not differ fundamentally from our data, except for the range of protrusion of reaction rate, which in our calculations is more powerful. Additionally, results of the reaction rate from Ref. 31 are shown by the blue squares. In addition, in Fig. 10, black rhombus and green triangles show reaction rates obtained in Refs. 14,46.

The small difference between these results and ours in the 0.1–3.0 $T_9$ region is apparently due to the fact that we are not always able to correctly describe *S*-factors, for transitions to some BSs. In the range of average energies calculated values turn out to be higher than experimental data. In Refs. 27,31,46,58, results of *R*-matrix parametrization of the *S*-factors were used to calculate reaction rate, which more accurately describes results of experimental measurements.

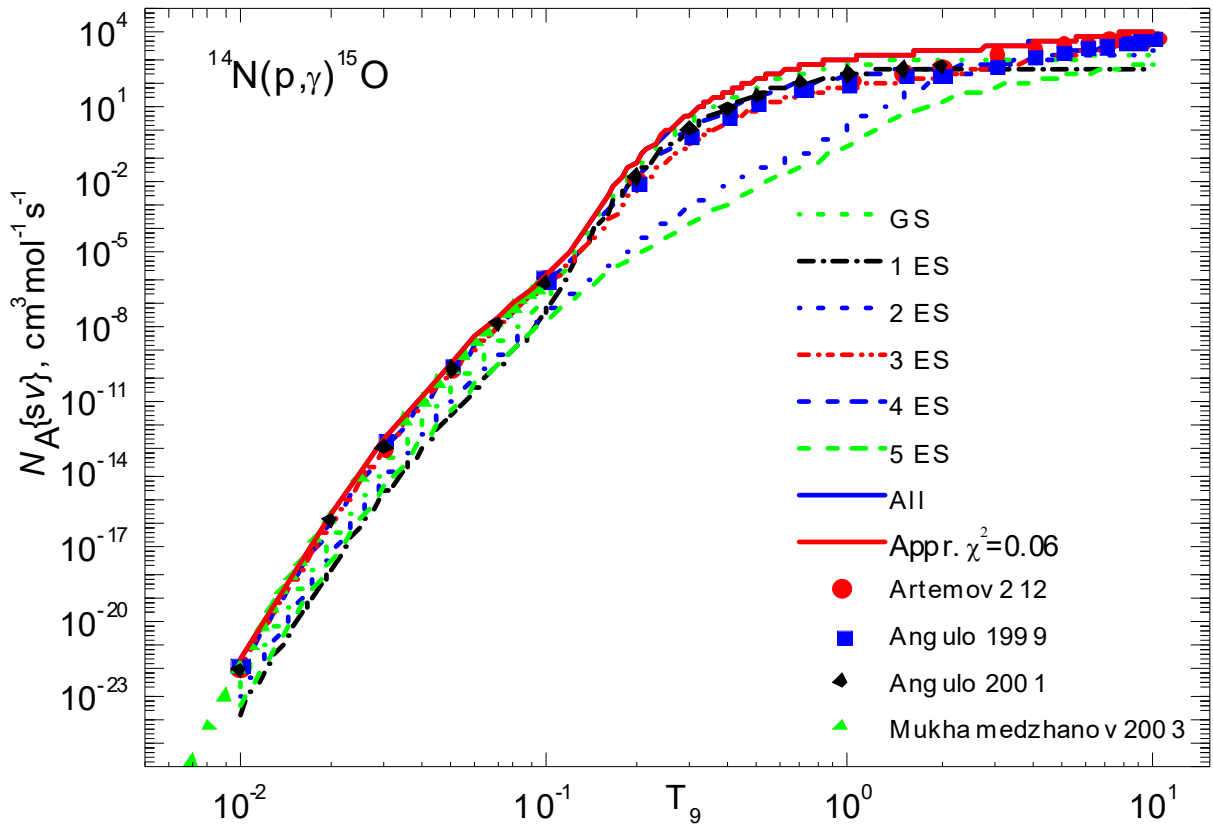

Fig. 10. Reaction rate of the radiative capture to all BS of $^{15}$O. Curves – results of our calculations, described in the text.

Furthermore, the parametrization of the obtained by us calculated reaction rate was carried out by the expression of the form Ref. 59

$$N_A \langle sv \rangle = \frac{a_1}{T^{2/3}} \exp(-\frac{a_2}{T^{1/3}})(1 + a_3 T^{1/3} + a_4 T^{2/3} + a_5 T + a_6 T^{4/3} + a_7 T^{5/3} + a_8 T^{7/3}) +$$

$$+ \frac{a_9}{T^{1/2}} \exp(-\frac{a_{10}}{T^{1/2}}) + \frac{a_{11}}{T} \exp(-\frac{a_{12}}{T}) + \frac{a_{13}}{T^{3/2}} \exp(-\frac{a_{14}}{T^{3/2}}) + \frac{a_{15}}{T^2} \exp(-\frac{a_{16}}{T^2})$$

Values of variation parameters are given in Table 17 (column No. 1 and No. 2), value $T$ denotes $T_9$, and the error of calculated reaction rate for calculating of $\chi^2$ was set at the level of 5%. Parameters for reaction rate from Table 17 lead to value of $\chi^2 = 0.06$, and results of calculations using this formula are shown in Fig. 10 with the red solid curve. This parametrization is practically coincide with the calculated reaction rate shown by the blue curve, and the blue curve cannot be seen in Fig. 10.

Table 17. Parameters of analytical parametrization of the reaction rate

| No. | $a_i$ | $a_i$ |
|---|---|---|
| 1 | 0.29960E+05 | 0.30210E+05 |
| 2 | 0.14538E+02 | 0.14536E+02 |
| 3 | 0.13897E+01 | 0.63266E+01 |
| 4 | 0.21168E+04 | 0.21168E+04 |
| 5 | -0.11164E+04 | -0.11164E+04 |
| 6 | -0.15854E+03 | -0.15847E+03 |
| 7 | 0.13177E+03 | 0.13178E+03 |
| 8 | -0.24814E+01 | -0.24986E+01 |
| 9 | 0.81765E+06 | 0.79902E+06 |
| 10 | 0.12109E+02 | 0.12109E+02 |
| 11 | 0.11707E+05 | 0.11689E+05 |
| 12 | 0.28095E+01 | 0.28098E+01 |
| 13 | 0.22053E+03 | 0.22197E+03 |
| 14 | 0.10896E+01 | 0.10896E+01 |
| 15 | 0.50490E+03 | 0.51004E+03 |
| 16 | 0.12177E+01 | 0.12212E+01 |
| | $\chi^2 = 0.06$ | $\chi^2 = 0.02$ (without two first points in Fig. 10) |

It can be seen that approximation describes calculation rate worse for the first two points, which are at the lowest temperatures. For all other points in $T_9$, magnitude of partial turns out to be less than one. If in such a parametrization the first two points are removed and a new minimization of $\chi_i^2$ is performed, its average value is approximately equal to 0.02 – these parameters are given in Table 17 in column No. 3.

## 6. Conclusion

As a result of this study of the astrophysical $S$-factor behavior in case of the proton radiative capture on $^{14}$N to five ES of $^{15}$O in the framework of two-body MPCM, acceptable description of available experimental data at energies up to 3–4 MeV is shown. In particular, it is possible to describe the astrophysical $S$-factors of the proton capture on $^{14}$N to these ES of $^{15}$O nucleus only under assumption that both scattering resonances at 260 keV and 987 keV are in the *D* waves. The use of *S* scattering waves for them does not lead to description of the cross section resonances and *S*-factors. For some bound states, for example, 4th ES is preferable to use the $D_{3/2}$ wave, for the 1st ES the $D_{1/2}$ wave, and for the 3rd ES $F_{3/2}$ wave. Improvement of description of available experimental data, using these states, is demonstrated by concrete calculations within the framework of MPCM.

The assumption made about presence of the $p^{14}N^*$ channel with excited $^{14}N^*$ cluster in the GS of $^{15}O$ allowed us to describe experimental data for capture to this BS, in general. It was possible to correctly describe not only the shape of the *S*-factor of such capture, but also order of its values. Note that this is the first case when in the framework of the MPCM it was necessary to use concept of an "excited cluster". Previously, on the basis of MPCM, we were able to describe the main characteristics of more than 30 reactions of radiative capture types like in Refs. 1,2,16–18,22,60,61 without going beyond basic states of clusters in the considered nuclei.

**Acknowledgments**

This work was supported by the Grant of Ministry of Education and Science of the Republic of Kazakhstan through the program BR05236322 "Study reactions of thermonuclear processes in extragalactic and galactic objects and their subsystems" in the frame of theme "Study of thermonuclear processes in stars and primordial nucleosynthesis" through the Fesenkov Astrophysical Institute of the National Center for Space Research and Technology of the Ministry of Defence and Aerospace Industry of the Republic of Kazakhstan (RK).

**References**

1. S. B. Dubovichenko, *Thermonuclear Processes in Stars and Universe* in: Second English edition, revised and expanded (Lambert Academy Publ., Saarbrucken, 2015); https://www.scholars-press.com/catalog/details/store/gb/book/978-3-639-76478-9/Thermonuclear-processes-in-stars.
2. S. B. Dubovichenko, *Radiative neutron capture and primordial nucleosynthesis of the Universe* in: Selected Methods for Nuclear Astrophysics, fifth Russian edition revised and expanded (Lambert Academy Publ., Saarbrucken, 2016); https://www.ljubljuknigi.ru/store/ru/book/radiative-neutron-capture/isbn/978-3-659-82490-6 (Accepted to press in de Gruyter in 2019).
3. V. G. Neudatchin *et al.*, *Phys. Rev. C* **45** 1512 (1992).
4. S. B. Dubovichenko and N. A. Burkova, *Russ. Phys. J.* **61** 852 (2018).
5. S. B. Dubovichenko and N. A. Burkova, *Russ. Phys. J.* **61** 1299 (2018).
6. S. B. Dubovichenko, N. A. Burkova, A. V. Dzhazairov-Kakhramanov and Ch. T. Omarov, *Russ. Phys. J.* **61** 1613 (2019).
7. S. B. Dubovichenko and N. A. Burkova, *Russ. Phys. J.* **61** 2105 (2019).
8. S. B. Dubovichenko, N. A. Burkova, A. V. Dzhazairov-Kakhramanov and T. A. Shmygaleva, *Izvestiya Vysshikh Uchebnykh Zavedenii, Fizika* **62** 22 (2019) (in Russian).
9. J. Grineviciute and D. Halderson, *J. Phys. G* **35** 055201(13p.) (2008).
10. Y. Xu., K. Takahashi, S. Goriely *et al.*, *Nucl. Phys. A* **918** 61 (2013).
11. J. T. Huang, C. A. Berutulani and V. Guimarães, *Atom Data Nucl Data* **96** 824-47

(2010).
12. Reza Ghasemi and Hossein Sadeghi, *Results in Physics* **9** 151-65 (2018).
13. A. Bohr and B. Mottelson *Nuclear structure* (Benjamin, New York, 1969).
14. C. Angulo and P. Descouvemont, *Nucl. Phys. A* **690** 755-68 (2001).
15. R. E. Azuma *et al*., *Phys. Rev. C* **81** 045805 (2010).
16. S. B. Dubovichenko and A. V. Dzhazairov-Kakhramanov, *Astrophys. J.* **819** 78(8p.) (2016).
17. S. B. Dubovichenko and A. V. Dzhazairov-Kakhramanov, *Nucl. Phys. A* **941** 335-63 (2015).
18. S. B. Dubovichenko, A. V. Dzhazairov-Kakhramanov and N. V. *Nucl. Phys. A* **963** 52-67 (2017).
19. S. B. Dubovichenko and Yu. N. Uzikov, *Phys. Part. Nucl.* **42** 251 (2011).
20. V. G. Neudatchin and Yu. F. Smirnov *Nucleon associations in light nuclei* (Nauka, Moscow, 1969).
21. H. Costantini, *et al*., *Rep. Prog. Phys*. **72** 086301(25p.) (2009).
22. S. B. Dubovichenko and A. V. Dzhazairov-Kakhramanov, *Int. Jour. Mod. Phys. E* **26** 1630009(56p.) (2017) and refs. to our papers in this journal.
23. C. Itzykson and M. Nauenberg, *Rev. Mod. Phys*. **38** 95 (1966).
24. S. B. Dubovichenko and A. V. Dzhazairov-Kakhramanov, *Russ. Phys. J.* **52** 833 (2009).
25. S. B. Dubovichenko, *Phys. Atom. Nucl.* **75** 173 (2012).
26. F. Ajzenberg-Selove, *Nucl. Phys. A* **523** 1 (1991); Revised manuscript Energy Levels of Light Nuclei *A*=15, 10 May 2002.
27. S. V. Artemov, *et al*., *Phys. Atom. Nucl.* **75** 291 (2012).
28. S. I. Sukhoruchkin and Z. N. Soroko, Excited nuclear states, Sub. G. Suppl. I/25 A-F, Springer, 2016.
29. W. A. Fowler, G. R. Caughlan and B. A. Zimmerman, *Ann. Rev. Astr. Astrophys*. **13** 69 (1975).
30. P. J. Mohr and B. N. Taylor, *Rev. Mod. Phys*. **77(1)** 1 (2005).
31. C. Angulo *et al*., *Nucl. Phys. A* **656** 3 (1999).
32. S. B. Dubovichenko and A. V. Dzhazairov-Kakhramanov, *Phys. Atom. Nucl.* **58** 579 (1995).
33. S. B. Dubovichenko and A. V. Dzhazairov-Kakhramanov, *Phys. Part. Nucl.* **28** 615 (1997).
34. J. M. Eisenberg and W. Greiner, *Excitation mechanisms of the nucleus* (North Holland, Amsterdam, 1976).
35. D. A. Varshalovich, A. N. Moskalev and V. K. Khersonskii, *Quantum Theory of Angular Momentum* (World Scientific, Singapore, 1989).
36. Constants in the category "Atomic and nuclear constants"; http://physics.nist.gov/cgi-bin/cuu/Value?mud|search_for=atomnuc!
37. V. V. Varlamov, B. S. Ishkhanov and S. Yu. Komarov, Chart of atomic nuclei; http://cdfe.sinp.msu.ru/services/ground/NuclChart_release.html.
38. M. P. Avotina and A. V. Zolotavin, *Moments of the ground and excited states of nuclei* (Atomizdat, Moscow, 1979).
39. P. E. Hodgson, *The Optical model of elastic scattering* (Clarendon Press, Oxford, 1963).
40. O. F. Nemets, V. G. Neudatchin, A. T. Rudchik, Y. F. Smirnov and Yu. M. Tchuvil'sky, *Nucleon association in atomic nuclei and the nuclear reactions of the* *many nucleons transfers* (Naukova dumka, Kiev, 1988).
41. Chart of nucleus shape and size parameters, 2015; http://cdfe.sinp.msu.ru/services/radchart/radmain.html
42. S. B. Dubovichenko, *Selected method for nuclear astrophysics* Second Russian Edition, corrected and enlarged (Lambert Academy Publ., Saarbrucken, 2012).
43. S. B. Dubovichenko, *Phys. Part. Nucl.* **44** 803 (2013).
44. G. R. Plattner and R. D. Viollier, *Nucl. Phys. A* **365** 8 (1981).
45. P. E. Bertone *et al.*, *Phys. Rev. C* **66** 055804 (2002).
46. A. M. Mukhamedzhanov *et al.*, *Phys. Rev. C* **67** 065804 (2003).
47. J. Bommer *et al.*, *Nucl. Phys. A* **172** 618 (1971).
48. W. Bohne *et al.*, *Nucl. Phys. A* **113** 97 (1968).
49. Q. Li *et al.*, *Phys. Rev. C* **93** 055806 (2016).
50. G. Imbriani, et al., *Euro. Phys. J. A* **25** 455 (2005).
51. U. Schroder *et al.*, *Nucl. Phys. A* **467** 240 (1987).
52. R. C. Runkle *et al.*, *Phys. Rev. Lett.* **94** 082503 (2005).
53. M. Marta *et al.*, *Phys. Rev. C* **83** 045804 (2011).
54. L. Wagner *et al.*, *Phys. Rev. C* **97** 015801 (2018).
55. E.G. Adelberger *et al.*, *Rev. Mod. Phys.* **83** 195 (2011).
56. A. Formicola *et al.*, *Nucl. Phys. A* **719** 94c (2003).
57. R. Depalo, *Int. Jour. Mod. Phys.: Conf. Ser.* **46** 1860003(8p.) (2018).
58. D. Bemmerer *et al.*, *Nucl. Phys. A* **779** 297 (2006).
59. G. R. Caughlan and W. A. Fowler, *Atom. Data Nucl. Data Tab.* **40** 283 (1988).
60. S. B. Dubovichenko and A. V. Dzhazairov-Kakhramanov, *Indian J. Phys.* **92(8)** 947 (2018); https://doi.org/10.1007/s12648-018-1190-8.
61. S. B. Dubovichenko, A. V. Dzhazairov-Kakhramanov and N. A. Burkova, *Int. J. Mod. Phys. E* **28** 1930004(49pp) (2019); doi: 10.1142/S0218301319300042.